\RequirePackage{fix-cm}
\documentclass[smallextended]{svjour3}       %
\smartqed  %

\usepackage{cite}
\usepackage{amsmath,amssymb,amsfonts}
\usepackage{algorithmic}
\usepackage{graphicx}
\usepackage{textcomp}
\usepackage[table,xcdraw]{xcolor}
\usepackage{tabularx}

\def\BibTeX{{\rm B\kern-.05em{\sc i\kern-.025em b}\kern-.08em
    T\kern-.1667em\lower.7ex\hbox{E}\kern-.125emX}}
    
\usepackage[hidelinks]{hyperref}
    
\usepackage{enumitem}

\usepackage{subcaption}

\usepackage{tcolorbox}

\usepackage{booktabs}
\usepackage{multirow}

\usepackage[normalem]{ulem}

\newcommand{\deleted}[1]{} 

\usepackage{xspace}
\newcommand{\tool}{\textsc{AmbiGuess}\xspace} %

\begin{document}

\title{Generating and Detecting True Ambiguity: A Forgotten Danger in DNN Supervision Testing\thanks{This work was partially supported by the H2020 project PRECRIME, funded under the ERC Advanced Grant 2017 Program (ERC Grant Agreement n. 787703).}
}

\author{Michael Weiss         \and
        André García Gómez      \and
        Paolo Tonella
}

\institute{Michael Weiss  \at
              Università della Svizzera italiana \\
              \email{michael.weiss@usi.ch}           %
           \and
           André García Gómez   \at
              Università della Svizzera italiana \\
              \email{andre.gg96@gmail.com}  
                         \and
           Paolo Tonella   \at
              Università della Svizzera italiana \\
              \email{paolo.tonella@usi.ch}  
}

\date{Received: date / Accepted: date}

\maketitle

\begin{abstract}

Deep Neural Networks (DNNs) are  becoming a crucial component of modern software systems, but they are prone to fail under conditions that are different from the ones observed during training (out-of-distribution inputs) or on inputs that are truly ambiguous, i.e., inputs that admit multiple classes with nonzero probability in their  labels.
Recent work  proposed DNN supervisors  to detect  high-uncertainty inputs before their possible misclassification leads to any harm.
To test and compare the capabilities of DNN supervisors, researchers proposed test generation techniques, to focus the testing effort on  high-uncertainty inputs that should be recognized as anomalous by supervisors. 
However, existing test generators aim to produce out-of-distribution inputs. No existing model- and supervisor independent technique targets the generation of truly \textit{ambiguous} test inputs, i.e., inputs that admit multiple classes according to expert human judgment.

In this paper, we propose a novel way to generate ambiguous inputs  to test DNN supervisors and used it to empirically compare several existing supervisor techniques.
In particular, we propose \tool to generate ambiguous samples for image classification problems. \tool is based on gradient-guided sampling in the latent space of a regularized adversarial autoencoder.
Moreover, we conducted what is -- to the best of our knowledge -- the most extensive comparative study of DNN supervisors, considering their capabilities to detect 4 distinct types of high-uncertainty inputs, including truly ambiguous ones. We find that the tested supervisors' capabilities are complementary: Those best suited to detect true ambiguity perform worse on invalid, out-of-distribution and adversarial inputs and vice-versa. 

\keywords{Neural Networks \and image classification \and ambiguity \and aleatoric uncertainty \and data-centric machine learning}

\end{abstract}

\section{Introduction}
\label{introduction}

Recently, more and more software systems are \emph{Deep Learning based Software Systems (DLS)}, i.e., they contain at least one  \emph{Deep Neural Network (DNN)}, as a consequence of the impressive performance that DNNs achieve in complex tasks, such as image, speech or natural language processing, in addition to the availability of affordable, but highly performant hardware (i.e., GPUs) where DNNs can be executed.
DNN algorithms can  identify, extract and interpret  relevant features in a training data set,  learning to  make  predictions about an unknown function of the inputs  at system runtime. 
Given the complexity of the tasks for which DNNs are used, predictions are typically made under uncertainty, where we distinguish between \emph{epistemic uncertainty}, i.e., model uncertainty which may be removed by better training of the model, possibly on better training data, and \emph{aleatoric uncertainty}, which is model-independent uncertainty, inherent in the prediction task (e.g., the prediction of a non-deterministic event). 
The former uncertainty is due to out-of-distribution (OOD) inputs, i.e., inputs that are inadequately represented in the training set.

The latter  may be due to ambiguity, i.e., an input for which multiple labels are all possibly correct (which could be understood as identical inputs having different, but correct, labels or -- more generally -- inputs having probabilistic labels).

This is a major issue often ignored during DNN testing, as recently recognized by Google AI Scientists: 
\emph{"many evaluation datasets contain items that (...) miss the natural ambiguity of real-world context"} \cite{Aroyo2021}.

The existence of uncertainty led to the development of \emph{DNN Supervisors} (in short, \emph{supervisors}), which aim to recognize inputs for which the DL component is likely to make incorrect predictions, allowing the DLS to take appropriate countermeasures to prevent harmful system misbehavior~\cite{Stocco2020, Henriksson2019, Henriksson2019a, Weiss2021, Weiss2022Stvr, Catak2021, hell2021monitoring, Hussain2022}.
For instance, the supervisor of a self-driving car might safely disengage the auto-pilot when detecting a high uncertainty driving scene~\cite{Stocco2020,wintersberger2021sdc}.
Other examples of application domains where supervision is crucial include medical diagnosis~\cite{davidson2021malaria, brown2021deep} and natural hazard risk assessment~\cite{bjarnadottir2019climate}.

While most recent literature on uncertainty driven DNN testing is focused on out of distribution detection~\cite{Henriksson2019, Henriksson2019a, Berend2020, Stocco2020, Zhang2018, Weiss2021, Weiss2022Stvr, Kim2018, Kim2021, Dola2021},
studies considering true ambiguity are lacking, which poses a big practical risk:
We cannot expect that supervisors which perform well in detecting epistemic uncertainty are guaranteed to perform well at detecting aleatoric uncertainty.
Actually, recent literature suggests the opposite~\cite{Mukhoti2021}.
The lack of studies considering true ambiguity is related to -- if not caused by -- the unavailability of ambiguous test data for common case studies: 
While to create ODD data, such as corrupted and adversarial inputs, a variety of precompiled dataset and generation techniques are publicly available ~\cite{Mu2019, Hendrycks2018, Rauber2017foolbox}, and invalid or mislabelled data is trivial to create in most cases, we are not aware of any approach targeting the generation of true ambiguity in a way that is sufficient for reliable and fair supervisor assessment.
In this paper we aim to close this gap by making the following contributions:

\begin{description}[noitemsep]
\item [Approach] We propose \tool, a novel approach to generate diverse, labelled, ambiguous images  for image classification tasks. Our approach is classifier independent, i.e., it aims to create data which is ambiguous to a hypothetical, perfectly well trained oracle (e.g., a human domain expert), and which does not just \emph{appear} ambiguous to a specific, suboptimally trained DNN. 
\item [Datasets] Using \tool, we generated and released two ready-to-use ambiguous datasets for common benchmarks in deep learning testing: MNIST~\cite{LeCun1998}, a collection of grayscale handwritten digits, and Fashion-MNIST~\cite{Xiao2017}, a more challenging classification task, consisting of grayscale fashion images.
\item [Supervisor Testing] Equipped with our datasets, we measured the capability of 16 supervisors at detecting different types of high-uncertainty inputs, including ambiguous ones.
Our results indicate that there is complementarity in the supervisors' capability to detect either ambiguity or corrupted inputs.
\end{description}

\section{Background} \label{sec:back}

\paragraph{Ambiguous Inputs} 
In many real-world applications, the data observed at prediction time might not be sufficient to make a certain prediction, even  assuming a hypothetical optimal oracle such as a domain expert with exhaustive knowledge: 
If some information required to make a correct prediction is missing, such missing information can be seen as a random influence, thus introducing aleatoric uncertainty in the prediction process.

Formally, in a given classification problem, i.e.,  a machine learning (ML) problem where the output is the class $c$ the input  $x$ is predicted to belong to, let $P(c \mid x)$ denote the probabilistic label, indicating the probability that $x$ belongs to $c$ in the ground truth's underlying distribution, where observation $x \in \mathbb{O}$ and $\mathbb{O}$ denotes the observable space, i.e., the set of all possibly observable inputs. 
We define \emph{true ambiguity} as follows:

\begin{definition}[True Ambiguity in Classification]
A data point $x \in \mathbb{O}$ is truly ambiguous if and only if $P(c \mid x) > 0 $ for more than one class $c$.
\end{definition}

Thus, inputs to a classification problem are considered \textit{truly ambiguous} if and only if such input is part of an overlap between two or more classes.
We emphasize \emph{true} ambiguity to indicate ambiguity intrinsic to the data and independent from any model and its classification confidence/accuracy.
In this way we distinguish ours from other papers which also use the term \emph{ambiguous} with different meaning, such as \emph{low confidence} inputs, \emph{mislabelled} inputs, where a label in the training/test set is clearly wrong, i.e, the corresponding probability in $P(c \mid x)$ is 0~\cite{Seca2021}, or \emph{invalid} inputs, where no true label exists for a given input\footnote{It can be noticed that the term invalidity is context dependent. 
Dola et. al.~\cite{Dola2021} consider an input invalid if it is out-of-distribution w.r.t. the training data, while still being an input which clearly belongs to one class, whereas other works consider as invalid input any relevant edge case~\cite{Mu2019, Hendrycks2018}.}. %
In simple domains, where humans may have no epistemic uncertainty (i.e., they know the matter perfectly), true ambiguity is equivalent to \emph{human ambiguity}.
In the remainder of this paper we focus only on true ambiguity and if not otherwise mentioned we use the term ambiguity as a synonym for \emph{true} ambiguity. %

\paragraph{Out-of-Distribution (OOD) Inputs} 
A prediction-time input is denoted OOD if it was insufficiently represented at training time, which caused the DNN  not to generalize well on such types of inputs. 
This is the primary cause of epistemic uncertainty.
OOD test data is used extensively to measure supervisor performance in academic studies, e.g.
by modifying nominal data in a model-independent, realistic and label-preserving way (\emph{corrupted} data)~\cite{Zhang2018, Mu2019, Hendrycks2018, Stocco2020}
or by minimally modifying nominal data to fool a specific, given model (\emph{adversarial} data).
In practice, both OOD and true ambiguity are important problems when building DLS supervisors \cite{Humbatova2020}.

\paragraph{Decision Frontier} Much recent literature works on the characterization of the decision frontier of a given model, i.e., its boundary of predictions between two classes in the input space~\cite{Karimi2019, Kang2020, Byun2020, Riccio2020Janus}.
It is imporant to note that the decision frontier is not equivalent to the sets of ambiguous inputs: 
The decision frontier is model specific, while ambiguity depends only on the problem definition and is thus independent of the model, i.e., the fact that an input is at a specific model's frontier, does not guarantee that it is indeed ambiguous (it may also be unambiguous, i.e., belong to a specific class with probability 1, or invalid, i.e., have 0 probability to belong to any class). 
The decision frontier may thus be considered the ``model's ambiguity'', while true ambiguity implies that an input is perceived as ambiguous by a hypothetical, perfectly well trained domain expert (hence matching ``human ambiguity'' in many classification tasks).

\section{Related Work}
\label{related}

\def \unctypesubfigurewidth {0.14\linewidth}
\def \unctypeimgwidth {.8\linewidth}

\begin{figure}
\centering
\begin{subfigure}{\unctypesubfigurewidth}
  \centering
  \includegraphics[width=\unctypeimgwidth]{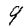}
  \caption{4/\underline{9}}
\end{subfigure}%
\begin{subfigure}{\unctypesubfigurewidth}
  \centering
  \includegraphics[width=\unctypeimgwidth]{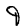}
  \caption{8/\underline{9}}
\end{subfigure}
\begin{subfigure}{\unctypesubfigurewidth}
  \centering
  \includegraphics[width=\unctypeimgwidth]{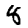}
  \caption{4/\underline{8}}
\end{subfigure}%
\begin{subfigure}{\unctypesubfigurewidth}
  \centering
  \includegraphics[width=\unctypeimgwidth]{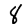}
  \caption{4/\underline{8}}
\end{subfigure}
\begin{subfigure}{\unctypesubfigurewidth}
  \centering
  \includegraphics[width=\unctypeimgwidth]{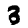}
  \caption{\underline{3}/8}
\end{subfigure}
\begin{subfigure}{\unctypesubfigurewidth}
  \centering
  \includegraphics[width=\unctypeimgwidth]{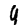}
  \caption{4/\underline{9}}
\end{subfigure}
\begin{subfigure}{\unctypesubfigurewidth}
  \centering
  \includegraphics[width=\unctypeimgwidth]{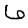}
  \caption{\underline{6}/9}
\end{subfigure}
\begin{subfigure}{\unctypesubfigurewidth}
  \centering
  \includegraphics[width=\unctypeimgwidth]{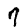}
  \caption{1/\underline{7}}
\end{subfigure}
\begin{subfigure}{\unctypesubfigurewidth}
  \centering
  \includegraphics[width=\unctypeimgwidth]{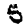}
  \caption{\underline{5}/9}
\end{subfigure}
\begin{subfigure}{\unctypesubfigurewidth}
  \centering
  \includegraphics[width=\unctypeimgwidth]{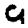}
  \caption{4/\underline{9}}
\end{subfigure}
\begin{subfigure}{\unctypesubfigurewidth}
  \centering
  \includegraphics[width=\unctypeimgwidth]{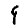}
  \caption{\underline{4}/9}
\end{subfigure}
\begin{subfigure}{\unctypesubfigurewidth}
  \centering
  \includegraphics[width=\unctypeimgwidth]{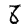}
  \caption{6/\underline{8}}
\end{subfigure}
\begin{subfigure}{\unctypesubfigurewidth}
  \centering
  \includegraphics[width=\unctypeimgwidth]{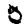}
  \caption{0/\underline{5}/9}
\end{subfigure}
\begin{subfigure}{\unctypesubfigurewidth}
  \centering
  \includegraphics[width=\unctypeimgwidth]{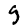}
  \caption{4/\underline{9}}
\end{subfigure}
\begin{subfigure}{\unctypesubfigurewidth}
  \centering
  \includegraphics[width=\unctypeimgwidth]{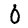}
  \caption{\underline{0}/8}
\end{subfigure}
\begin{subfigure}{\unctypesubfigurewidth}
  \centering
  \includegraphics[width=\unctypeimgwidth]{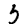}
  \caption{\underline{3}/5}
\end{subfigure}
\begin{subfigure}{\unctypesubfigurewidth}
  \centering
  \includegraphics[width=\unctypeimgwidth]{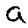}
  \caption{0/\underline{9}}
\end{subfigure}
\begin{subfigure}{\unctypesubfigurewidth}
  \centering
  \includegraphics[width=\unctypeimgwidth]{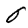}
  \caption{0/\underline{6}}
\end{subfigure}

\caption{
The 18 most ambiguous images, manually selected from the 300 (3\%) samples with the highest predictive entropy in the MNIST test set~\cite{LeCun1998}. Only a few of them are clearly ambiguous, showing that ambiguous data are scarce in existing datasets. 
Underlined numbers show the actual label, non-underlined numbers show classes 
we consider possibly having a non-zero probability as well (making the image ambiguous).
}
\label{fig:mnist-nominal-amb}
\end{figure}

The research works that are most related to our approach deal with automated test generation for DNNs~\cite{Mu2019,Hendrycks2018,Tian2018Deeptest,Zhang2018,Stocco2020,Rauber2017foolbox}.
In these works, some reasons for uncertainty, such as ambiguity, are not considered. Hence, automatically generated tests do not allow meaningful evaluations under ambiguity of DNN supervisors, as well as of the DNN behavior, in the absence of supervisors.
We illustrate this in \autoref{fig:mnist-nominal-amb}:
Using an off-the-shelf MNIST~\cite{LeCun1998} classifier, we calculated the predictive entropy to identify the 3\% of samples (300 out of 10'000) with the presumably highest aleatoric uncertainty in the MNIST test set.  Predictive entropy (i.e., the entropy of the Softmax values interpreted as probabilities) is a standard metric used in the related literature~\cite{Mukhoti2021} to detect aleatoric uncertainty which is caused, amongst other reasons, by truly ambiguous images.
Out of these 300 images, we manually selected the ones we considered potentially ambiguous, and show them in \autoref{fig:mnist-nominal-amb}.
Clearly, some of them are ambiguous, showing that ambiguity exists and is present in the MNIST test set, but the scarcity of truly ambiguous inputs indicates that supervisors  cannot be confidently tested for their capability of handling ambiguity using this test set.
The manual selection of the 18 (subjectively) most ambiguous images was required to exclude the 282 images that also had high entropy, but did not appear truly ambiguous: For some of them, the high entropy was clearly caused by image invalidity. For others, the high entropy was caused by the model's inability to assign a high likelihood to a single class for an unambiguous, nominal image. 
The latter serves as an example showing that using just the Softmax value to detect ambiguity might not be ideal and highlights the need for an empirical comparison of the different supervisors' capability to detect ambiguity (see \autoref{benchmark}).

In the DNN \emph{test input generators} (TIG) literature~\cite{Mu2019, Hendrycks2018, Tian2018Deeptest, Zhang2018, Stocco2020, dunn2021exposing},
with just one notable preprint as an exception~\cite{Mukhoti2021}, we are not aware of any paper aiming to generate true ambiguity directly, while most TIG aim for other objectives.
Some works~\cite{Mu2019, Hendrycks2018, weiss2022simpleTip} propose to corrupt nominal input in predefined, natural and label-preserving ways to generate OOD test data.
DeepTest~\cite{Tian2018Deeptest} applies corruptions to road images, e.g., by adding rain, while aiming to generate data that maximizes neuron coverage.
Also targeting road images, DeepRoad~\cite{Zhang2018} is a framework using Generative Adversarial Networks  (GAN) to change conditions (such as the presence of snow) on nominal images.
The Udacity Simulator, used by Stocco et. al., \cite{Stocco2020}, allows to dynamically add corruptions, such as rain or snow, when testing self-driving cars.
Similar to DeepTest, TensorFuzz~\cite{Xie2019} and DeepHunter~\cite{Odena2019} generate data with the objective to increase test coverage. 
Again, aiming to generate diverse and unseen inputs, these approaches will mostly generate OOD inputs and only occasionally -- if at all -- truly ambiguous data. 

A fundamentally different objective is taken in adversarial input generation~\cite{Goodfellow2014explainingAdvFGSM}, where nominal data is not changed in a natural, but in a malicious way. 
Based on the tested model, nominal input data is slightly changed to cause misclassifications.
Literature and open source tools provide access to a wide range of different specific adversarial attacks~\cite{Rauber2017foolbox}.
While very popular, neither input corruptions nor adversarial attacks generate intentionally ambiguous data from nominal, typically non-ambiguous inputs. As they rely on the ground truth label of the modified input to remain unchanged, they do not aim at creating true ambiguity, as affecting the ground truth label would imply unsuccessful test data generation.

Another popular type of test data generators aims to create inputs along the decision boundary:
DeepJanus~\cite{Riccio2020Janus} uses a model based approach, while SINVAD~\cite{Kang2020} and MANIFOLD~\cite{Byun2020} use the generative power of variational autoencoders (VAE)~\cite{Kingma2013}.
Note that we cannot expect inputs along the decision boundary to be always truly ambiguous 
-- they may just as well be OOD, invalid or in rare cases even low-uncertainty inputs. 
In addition, these approaches are by design model specific, making them unsuitable to generate a generally applicable, model-independent, ambiguous dataset.

Thus, out of all the approaches discussed above, none aims to generate a truly ambiguous dataset. 
A notable exception is a recent, yet unpublished, preprint by Mukhoti et. al., \cite{Mukhoti2021}. 
In their work, to evaluate the uncertainty quantification approach they propose, they needed an ambiguous MNIST dataset.
To that extent, they used a VAE to generate a vast amount of data (which also contains invalid, OOD and un-ambiguous data) which they then filter and stratify based on two mis-classification prediction techniques, aiming to end up with a dataset consisting of ambiguous images. 
We argue that, while certainly valuable in the scope of their paper, the so-created dataset is not sufficient as a standard benchmark for DNN supervisors, as the approach itself relies on a supervision technique, hence being circular if used for DNN supervisor assessment. In fact, the created ambiguity may  be particularly hard (or easy) to be detected by supervisors using different (resp. similar) MP techniques. 
We anyway compared their approach to ours empirically and found that it is less successful in generating truly ambiguous test data than ours.

\section{Uses of Ambiguous Test Sets}
\label{motivation}

In this paper we focus on the usage of ambiguous test data for the assessment of DNN supervisors, but ambiguous data have also other uses, including the assessment of test input prioritizers.

\subsection{Assessment of DNN supervisors}
We cannot assume that results on DNN supervisors' capabilities obtained on nominal and OOD data generalize to ambiguous data.
Recent studies~\cite{Zhang2020, Weiss2021, Weiss2022Stvr} have shown that there is no clear performance dominance amongst uncertainty quantifiers used as DNN supervisors, but such studies overlook the threats possibly associated with  the presence of ambiguity. Warnings on such threats  in medical machine learning based systems were raised already in 2000 \cite{trappenberg2000classification}, with ambiguity in a cancer detection dataset mentioned as a specific example. The authors proposed to equip the system with an ambiguity-specific supervisor, to ``detect and re-classify as ambiguous''~\cite{trappenberg2000classification} such threatening data. 
To test such supervisors, such as the one proposed by Mukhoti et. al.~\cite{Mukhoti2021}, model and MP independent and diverse ambiguous data is needed. 

\subsection{Assessment of DNN input prioritizers} Test input prioritizers, possibly based on MP, aim to prioritize test cases (inputs) in order to allow developers to detect mis-behaviours (e.g., mis-classifications) as early as possible. Hence, they should be able to recognize ambiguous inputs. Correspondingly, test input prioritizers should   be assessed also on ambiguous inputs.
On the contrary, when the goal is active learning, an ambiguous input should  be given the least priority or excluded at all, as the aleatoric uncertainty causing its mis-classification cannot by definition be avoided using more training data.
Thus, recognition of ambiguous test data is clearly of high importance when developing a test input prioritizer, be that to make sure that the ambiguous samples are given a high priority (during testing) or a low priority (during active learning).

\subsection{Decision-Boundary Oracle}
\label{sec:applications_dec_boundary_oracle}
\begin{figure} 
\centering
\fbox{\includegraphics[width=0.5\textwidth]{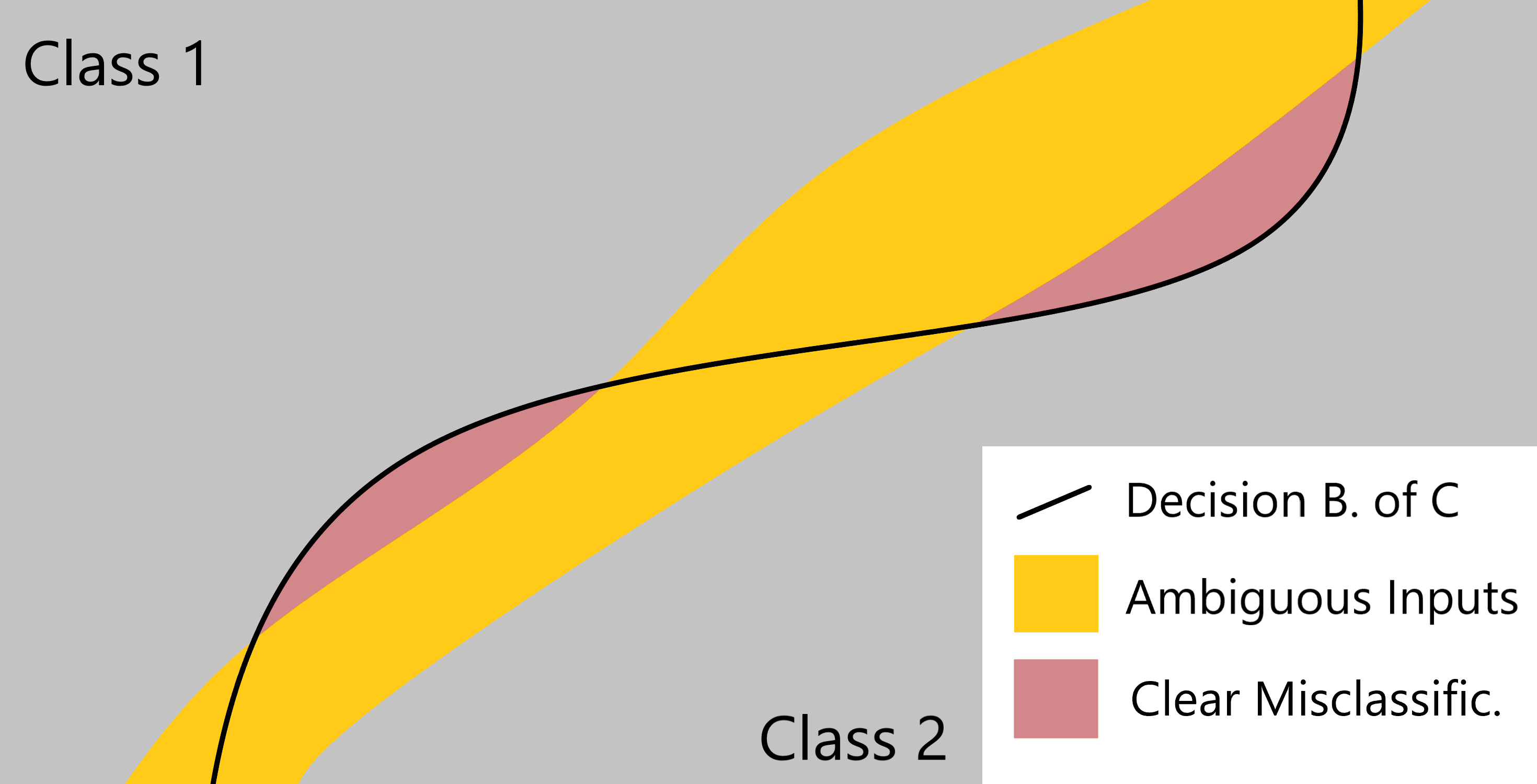}}
\caption{Schematic segmentation of a valid input space:
If two classes are separated by ambiguous inputs, a decision boundary of classifier C outside of these ambiguous inputs implies (unambiguous) misclassifications.}
\label{fig:amb_vs_db}
\end{figure}

Much recent literature works on the characterization of the decision boundary of a given model, i.e., its frontier of predictions between two classes in the input space~\cite{Karimi2019, Kang2020, Byun2020, Riccio2020Janus}.
Given that for an ambiguous sample, two or more classes can be considered as true labels, we would expect all ambiguous samples to lie close to the decision boundary of a well trained classifier.
Similarly, considering only the \emph{valid input space}, i.e., the subset of the input space which contains the valid inputs for the given classification problem, the presence of an \emph{ambiguous space (AS)}, i.e., of truly ambiguous samples, implies that the decision boundary of said classifier must go through the AS. 
This is illustrated in \autoref{fig:amb_vs_db}, which shows an ambiguous space and the decision boundary of a suboptimally trained classifier $C$. 
The fact that the decision boundary is not always within the AS implies that the inputs lying between the decision boundary and the AS consist of unambiguous samples misclassified by $C$.
Moreover we know that adding data from these enclosed (clear misclassification) areas to the training set will increase the performance of $C$, which is not necessarily true for samples within the AS. 
Hence, knowledge about the ambiguity of data near the decision boundary is important to assess the quality of a model and possibly to improve it, when unambiguous data is found at the frontier.

It should be noticed that in this paper, we do not make any assumptions about the decision boundary and its connection to truly ambiguous inputs and \autoref{fig:amb_vs_db} serves only as illustration of a situation that may occur in practice.
Indeed, in \autoref{benchmark}, we compare  supervisors directly relying on the predicted probabilities and thus also on the decision boundary (such as Vanilla Softmax) with some that do not (such as autoencoders ~\cite{Stocco2020}). Indeed, the good results of Vanilla Softmax and other techniques relying on the decision boundary do suggest that ambiguous samples are very likely to lie close to such boundary.

\subsection{Disentanglement and Reasoning}

The identification of ambiguity can be seen as a special case of \textit{uncertainty disentanglement and reasoning}~\cite{Lines2019Disentangling, Clements2019}, the former being the quantification of epistemic vs aleatoric uncertainty, and the latter being the separation of uncertainty into specific root causes, such as data invalidity, OOD, or true ambiguity.
Recent work has used uncertainty disentanglement to guide training in reinforcement learning~\cite{Lines2019Disentangling, Clements2019},
building on the idea that only data leading to epistemic uncertainty is useful to drive model performance improvement during continuous learning tasks. 
Let us consider the following example:

\begin{example}[Use of Uncertainty Reasoning] A medical DLS determines if a patient has a specific type of cancer, provided some ultrasonic images.
\begin{enumerate}
    \item[(1)] Assume the ultrasonic image reveals an implant of the patient -- something which is underrepresented in the training set, making the input OOD and potentially leading the DNN to mistake the implant as something relevant for cancer detection: 
    Being able to reliably detect that the input is OOD, the system could ask a (human) expert to label the image. Said label would then be a more reliable prediction, as the human is not confused by the implant. In addition, the now labelled OOD sample can be used in further training loops of the DNN.
    \item[(2)] Assume the input is in-distribution, but there's not enough information on the image to decide if the patient has cancer: The image is truly ambiguous. By recognizing this true ambiguity, the DLS may make a reliable probabilistic prediction, which would allow the patient to make an informed decision on whether to conduct further diagnosis or treatment.
    \item[(3)] Assume the DNN is given an image which is not an ultrasonic image. Detecting that this input is invalid allows the system to refuse to make any (even probabilistic) prediction and raise an alert. %
\end{enumerate}
Case (2) is a particularly realistic case: In AI-guided healthcare, decisions about future treatment and diagnosis  are typically made based on probabilistic predictions~\cite{DeHond2022HealthcareAiGuidelines}, which can only be trusted if the input is in-distribution.
\end{example}

Another reason for fine-grained uncertainty reasoning is DLS debugging: 
Informing the developers of a DLS about the root causes of uncertainties and mispredictions would greatly facilitate further improvement of the DLS, especially because DNNs are known for their low explainability~\cite{Samek2017}, which makes debugging particularly challenging when dealing with them.
Clearly, to develop and test any technique working with uncertainty disentanglement or uncertainty reasoning, the availability of ambiguous data in the test set is a strict prerequisite, and the lack of such datasets is likely the main reason why  such research is so scarce.
\section{Generating Ambiguous Test Data}
\label{generation}
We designed \tool, a TIG targeting ambiguous data for image classification, based on the following design goals (DG):

\noindent
\textbf{DG\textsubscript{1} (labelled ambiguity):}
The generated data should be truly ambiguous and have correspondingly \textit{probabilistic labels}, i.e., each generated data is associated with a probability distribution over the set of labels.
Probabilistic labels are the most expressive description of true ambiguity and a single or multi-class label can be trivially derived from probabilistic labels.

\noindent
\textbf{DG\textsubscript{2} (model independence):} 
To allow universal applicability of the generated dataset, our TIG should not depend on any specific DNN under test.

\noindent
\textbf{DG\textsubscript{3} (MP independence):} 
The created dataset should allow fair comparison between different supervisors. Since supervisors are often based on MPs (e.g., uncertainty or confidence quantifiers), our TIG should not use any MP as part of the data generation process, to avoid circularity, which might give some supervisor an unfair advantage or disadvantage over another one. 

\noindent
\textbf{DG\textsubscript{4} (diversity):}
The approach should be able to generate a high number of diverse images.

\subsection{Interpolation in Autoencoders}
\begin{figure}
    \centering
    \includegraphics[width=\linewidth]{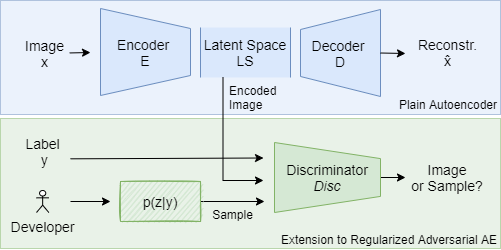}
    \caption{Autoencoder (blue) and its extension to a Regularized Adversarial Autoencoder (green)}
    \label{fig:ae-and-raae}
\end{figure}

Autoencoders (AEs)
are a powerful tool, used in a range of TIG~\cite{Kang2020, Byun2020, Mukhoti2021, dunn2021exposing}.
AEs follow an \emph{encoder-decoder} architecture as shown in the blue part of \autoref{fig:ae-and-raae}: 
An encoder $E$ compresses an input into a smaller \emph{latent space (LS)}, and the decoder $D$ then attempts to reconstruct $x$ from the LS. 
The reconstruction loss, i.e., the difference between input $x$ and reconstruction $\hat{x}$ is used as the loss to be minimized during training of the AE.

\begin{figure} 
\centering
\includegraphics[trim=0 0 500 95, clip, width=5.2cm]{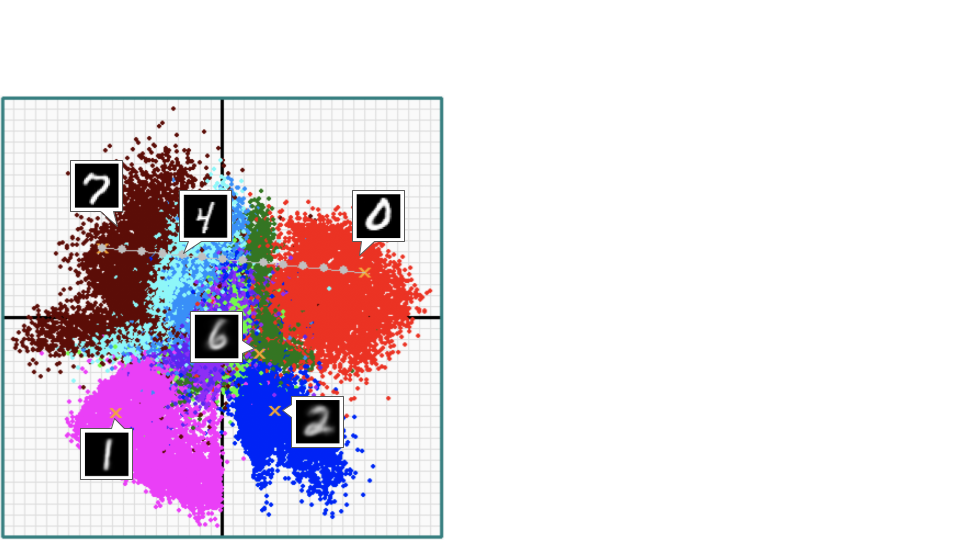}
\caption{Image Sampling in the Latent Space}
\label{fig:ls_sampling}
\end{figure}

On a trained AE, sampling arbitrary points in the latent space, and using the decoder to construct a corresponding image, allows for cheap image generation.
This is shown in \autoref{fig:ls_sampling}, where the shown images are not part of the training data, being reconstructions based on randomly sampled points in the latent space.
In the following section, we leverage the generative capability of AEs, by proposing an architecture that can target ambiguous samples specifically and can label the generated data probabilistically (\textit{DG\textsubscript{1}}).

\subsection{\tool}
Our TIG \tool consists of three components: 
(1) The \emph{Regularized LS Generation} component, which trains a specifically designed AE to have a LS that facilitates the generation of truly ambiguous samples. 
(2) The \emph{Automatic Labelling} component, which leverages the AE architecture to support probabilistic labelling of any images produced by the AE's decoder.
(3) The \emph{Heterogenous Sampling} component, which chooses samples in the LS in a way that leads to high diversity of the generated images.

\begin{figure}
    \centering
    \includegraphics[width=\linewidth]{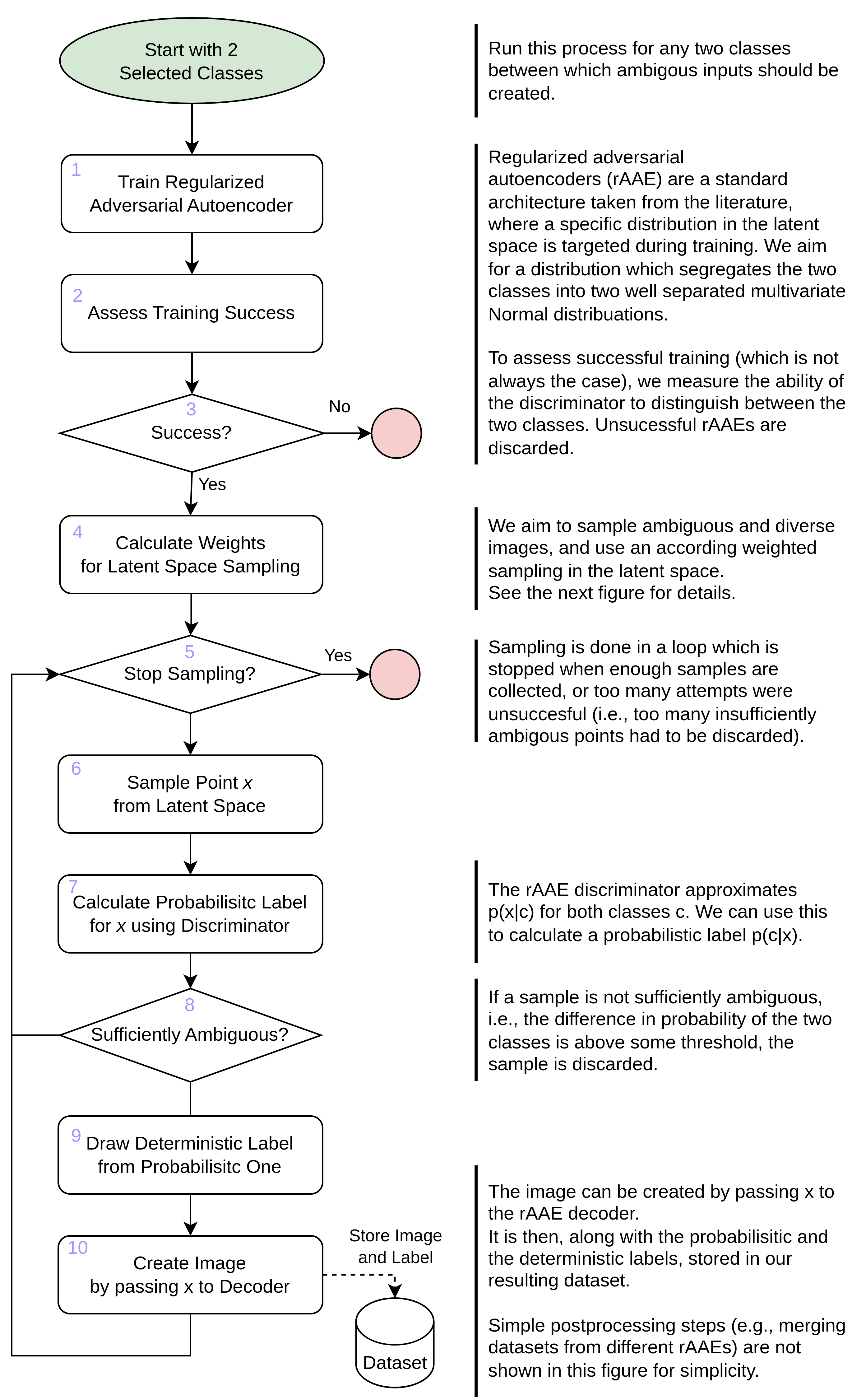}
    \caption{High-Level Illustration of \tool}
    \label{fig:ambiguess-outline}
\end{figure}

An overview of \tool, which leverages these three components, is outlined in \autoref{fig:ambiguess-outline}.

\subsubsection{Regularized Latent Space Generation}
\label{sec:reg-lat-space}

Interpolation from one class to another in the latent space, i.e., the gradual perturbation of the reconstruction by moving from one cluster of latent space points to another one, may produce ambiguous samples between those two classes (satisfying both \textit{DG\textsubscript{2}} and \textit{DG\textsubscript{3}}). 
An example of such an interpolation is shown in \autoref{fig:interpolation}.
Clearly, we want the two clusters to be far from each other, providing a wide range for sampling in between them, and no other cluster should be in proximity, as it would otherwise influence the interpolation. However, these two conditions are usually not met by traditional autoencoders used in other TIG approaches. 
For example, \autoref{fig:ls_sampling} shows the LS of a standard variational autoencoder (a popular architecture in TIG).
Here, interpolating between classes 0 and 7 would, amongst others, cross the cluster representing class 4, and thus samples taken from the interpolation line would clearly not be ambiguous between 0 and 7, but would be reconstructed as a 4 (or any of the other clusters lying between them). 
We solve these requirements by using 2-class Regularized Adversarial AEs:

\begin{figure} 
\centering
\includegraphics[trim=100 45 100 85, clip, width=8cm]{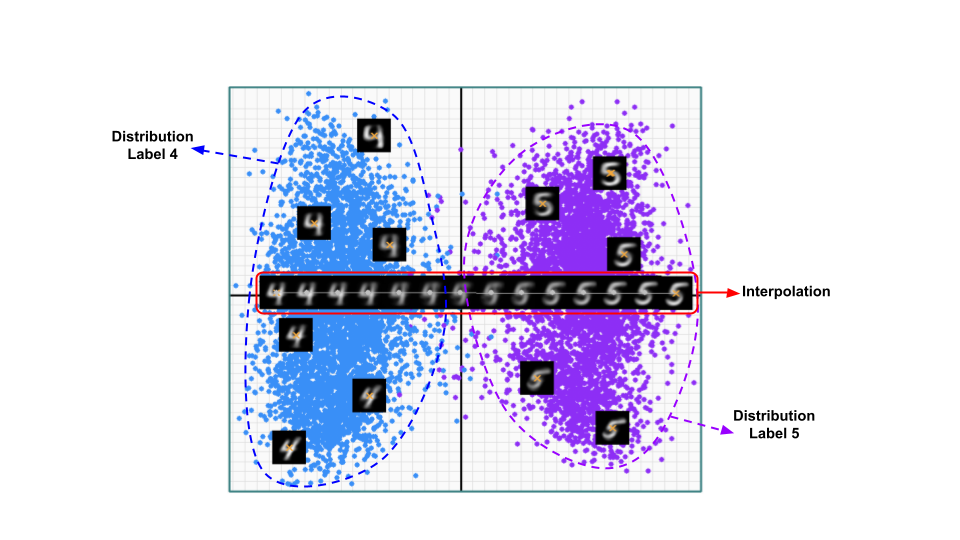}
\caption{Interpolation between two classes in the latent space of a 2-class Regularized Adversarial Autoencoder.}
\label{fig:interpolation}
\end{figure}

\paragraph{2-class AE}~Instead of training one AE on  all classes, we train multiple AEs, each one with the training data of just two classes. 
This has a range of advantages: First and foremost, it prevents interferences with third classes. Then, as the corresponding reduced (2-class) datasets have naturally a lower variability (feature density), 2-class autoencoders are expected to require fewer parameters and show faster convergence during training.
Further, the fact that the number of combinations of classes ${c}\choose{2}$ grows exponentially in the number of classes $c$ is of only limited practical relevance: In very large, real-world datasets, ambiguity is much more prevalent  between some combinations of classes than between others, so not all pairwise combinations are equally interesting for the test generation task.
For example, let us consider a self driving car component which classifies vehicles on the road. While an image of a vehicle where one cannot say for sure weather it is a pick-up or a SUV (hence having true ambiguity) is clearly a realistic case, an image which is truly ambiguous between a SUV and a bicycle is hard to imagine. 
This phenomenon is well known in the literature, as it leads to \emph{heteroscedastic} aleatoric uncertainty~\cite{Ayhan2018}, i.e., aleatoric uncertainty which is more prevalent amongst some classes than amongst others.
In such a case, using \tool, one would only construct the 2-class AEs for selected combinations where ambiguity is realistic.

\paragraph{Regularized Adversarial AE (rAAE)}
To guide the training process towards creating  two disjoint clusters representing the two classes, with an adequate amount of space between them, we use a \emph{Regularized Adversarial Autoencoder (rAAE)}~\cite{Makhzani2015}.
The architecture of an rAAE is shown in \autoref{fig:ae-and-raae}:
Encoder $E$, Decoder $D$ and the LS are those of a standard AE. 
In addition, similar to other adversarial models~\cite{Goodfellow2014}, a discriminator $Disc$ is trained to distinguish labelled, encoded images $z$ from samples drawn from a predefined distribution $p(z|y)$.
Specifically, we define $p(z|y)$ as a multi-modal (2 classes) multi-variate (number of dimensions in latent space) gaussian distribution, consisting of $p(z|c_1)$ and $p(z|c_2)$ for classes $c_1$ and $c_2$, respectively. 
Then, training a rAAE consists of three training steps, which are executed on every training epoch:
First, similar to a plain AE, $E$ and $D$ are trained to reduce the reconstruction loss. 
Second, $Disc$ is trained to discriminate encoded images from samples drawn from $p(z|y)$,
and third, $E$ is trained to fool $Disc$, i.e., $E$ is trained with the objective that the training set projected onto the latent space matches the distribution $p(z|y)$.
This last property can be leveraged for  ambiguous test generation:
Given two classes $c_1$ and $c_2$, to clear up space between them in the latent space we can choose a $p(z|y)$ such that  $p(z|c_1) > \epsilon$ on LS points disjoint from the LS points where $p(z|c_2) > \epsilon$,  for some small $\epsilon$. 
For example, assume a two-dimensional latent space: 
Choosing $p(z|c_1) = \mathcal{N}([-3,0], [1,1])$ and $p(z|c_2) = \mathcal{N}([3,0], [1,1])$ will, after successful training, lead to a latent space where points representing $c_1$ are clustered around $(-3,0)$ and points representing $c_2$ around $(3,0)$, with few if any points between them, i.e., around $(0,0)$. This makes reconstructions around $(0,0)$ potentially highly ambiguous.

\subsubsection{Probabilistic Labelling of Images}
\label{sec:prob-labelling}
The $Disc$ of a 2-class rAAE can be used to automatically label the images generated by its decoder:
Given a latent space sample $z^*$ on a 2-class rAAE for classes $c_1$ and $c_2$, $Disc(z^*, c_1)$ approximates $p(z^*|c_1)$. Assuming $p(c_1) = p(c_2) = 0.5$, we have $p(z^*|c_1) = p(c_1 | z^*)$. Hence, $Disc(z^*, c_1)$ approximates the likelihood that $z^*$ belongs to class $c_1$. The same holds for $Disc(z^*, c_2)$.
Normalizing these two values s.t. they add up to  1 thus provides a probability distribution over the classes (thus realizing \textit{DG\textsubscript{1}}). This is used in Steps 4 and 7 of \autoref{fig:ambiguess-outline}.

Clearly, this probabilistic labelling depends on the discriminator being well trained, i.e., its ability to discriminate between images of classes $c_1$ and $c_2$. Thus, we propose to assess the discriminator's training success by measuring its accuracy at classifying nominal (non-ambigous) inputs as one of the two classes. Then, rAAEs for which the discriminators accuracy does not meet a (tunable) threshold can be discarded (Steps 3 and 4 in \autoref{fig:ambiguess-outline}).

\subsubsection{Selecting Diverse Samples in the LS}
\begin{figure}
    \centering
    \includegraphics[width=\linewidth]{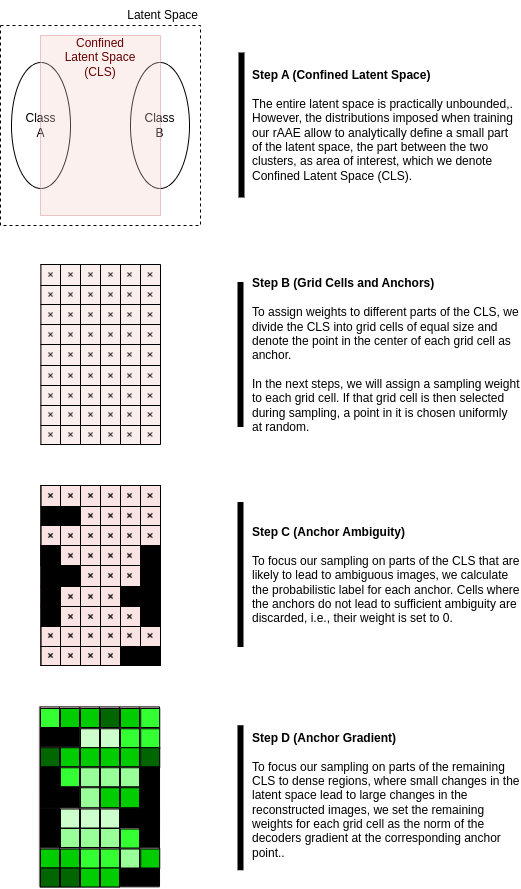}
    \caption{Illustration of the Weights-Calculation in the Latent Space}
    \label{fig:sampling-weigths-illustration}
\end{figure}
Diversity in a generated dataset (see \textit{DG\textsubscript{4}}) is in general hard to achieve when generating a dataset by sampling  the LS, as the distance between two points in the LS does not directly translate to a corresponding difference between the  generated images.
While in some parts of the LS, which we denote as \emph{high density} parts, moving a point slightly in the LS space can lead to clearly visible changes in the decoder's output, in \emph{low density} parts, large junks of the LS lead to very similar reconstructed images.

To that extent, we do not sample in the latent space uniformly, but in a weighted way aiming to select diverse images (Step 4 in \autoref{fig:ambiguess-outline}). Specifically, we set up the sampling of points in the latent space in four steps as outlined below, also illustrated in~\autoref{fig:sampling-weigths-illustration}:
\begin{enumerate}
    \item[A.] \textbf{Confined Latent Space} The size of latent space is practically infinite, being bound only by its numerical representation. However, we are only interested in a small part of this latent space, namely the area in between the two 'nominal' clusters in our 2-class rAAE.
    This is represented in Step A of \autoref{fig:sampling-weigths-illustration}. 
    This area, which we denote as \emph{confined latent space (CLS)}, can be defined analytically from the distributions imposed on the 2-class rAAE during training.\footnote{
    Specifically, in the datasets generated in this paper, we use two-dimensional latent spaces. The distributions are defined similar to the illustrations in \autoref{fig:sampling-weigths-illustration} as two multivariate normal distributions next to each other. On the first (horizontal) axis, the CLS is thus constrained by the means of the distributions on this axis. On the second (vertical) axis, both clusters have the same mean and we use +- 5 standard deviations from that mean as bounds of the CLS.}
    In the subsequent steps, we consider only the CLS. 
    \item[B.] \textbf{Grid Cells and Anchors} We divide the (rectangular) CLS into a grid of rectangular grid cells, where the number of grid cells is a tunable hyperparameter. Then, for every grid cell we identify the point in the center. In the next two steps, we will use this anchor point as a representative of the grid cell when estimating the density in the grid cell, as well as the ambiguity in the images reconstructed from points within the grid cell.
    With this, we will build a weight for each cell, based on which cells are selected during sampling. Within a grid cell, the actually drawn point will then be chosen uniformly at random.
    \item[C.] \textbf{Anchor Ambiguity} We calculate the probabilistic label for each anchor, as explained in Section \ref{sec:prob-labelling}. For labels which are not sufficiently ambiguous, i.e., where the  difference between the two class probabilities is higher than some threshold $\delta_{max}$, the corresponding grid cells are ignored (their sampling weight is set to zero). Thus, $\delta_{max}$ is a hyperparameter allowing us to steer the minimum level of ambiguity in the anchors of the cells used for sampling. Note that, as points in the grid cells exhibit lower ambiguity than the corresponding anchor, $\delta_{max}$ does not aim to ensure this level of ambiguity in the resulting test set; this is ensured with a final filtering (Steps 7 and 8 in \autoref{fig:ambiguess-outline}). However, $\delta_{max}$ enables the sampling algorithm to consider only regions (i.e., grid cells) of the CLS with a high likelihood to generate ambiguous images, making it overall much more efficient.
    \item[D.] \textbf{Anchor Gradient} We want to focus our sampling on high density regions of the latent space where small changes in the latent space representation lead to more notable changes in the reconstructed images than in low density regions. We thus estimate the density of each none-ignored grid cell by calculating the norm of the decoders gradient at the corresponding anchor point. We use the euclidean distance to measure differences between decoder outputs (i.e., images), which is required to calculate the gradient.
    We then use these density estimates (i.e., these norms) as weights when choosing grid cells during sampling.
\end{enumerate}

\subsection{Pre-Generated Ambiguous Datasets}
\label{sec:generated_datasets}
We built and released two ready-to-use ambiguous datasets for  \emph{MNIST}~\cite{LeCun1998}, the most common dataset used in software testing literature~\cite{Riccio2020}, where images of handwritten numbers between 0 and 9 are to be classified, and its more challenging drop-in replacement \emph{Fashion MNIST (FMNIST)}~\cite{Xiao2017}, consisting of images of 10 different types of fashion items.

\paragraph{\tool configuration} 
For each pair of classes, we trained 20 rAAEs to exploit the non-determinism of the training process to generate even more diversified outputs.
To make sure we only use rAAEs where the distribution in the LS is as expected, we check if the discriminator cannot distinguish LS samples obtained from input images w.r.t. LS samples drawn from $p(z|y)$: the accuracy on this task should be between $0.4$ and $0.6$. At the same time, we check if the discriminator's accuracy in assigning a higher probability to the correct label of nominal samples is above 0.9. Otherwise it is discarded. 
Combined, we used the resulting rAAEs to draw 20,000 training and 10,000 test samples for both MNIST and FMNIST, using $\delta_{max}=.25$ for test data and  $\delta_{max}=0.4$ for training data. We ignored generated samples where the difference between the two label's probabilities was above  $\delta_{max}$.
We chose different $\delta_{max}$, (the loose upper threshold of difference in the two class probabilities) for train and test set as our test set should be clearly and highly ambiguous, e.g. to allow studies that specifically target ambiguity (hence a low $\delta_{max}$). In turn, the training set should more continuously integrate with the nominal data, hence we also allow  for less ambiguous data compared to our ambiguous test set.

\section{Evaluation of Generated Data}
\label{evaluation}

The \textbf{goal} of this experimental evaluation is \textit{to assess both quantitatively and qualitatively whether \tool can indeed generate truly ambiguous data}.
We evaluate the ambiguity in our generated datasets first using a quantitative analysis where we analyze the outputs of a standard, well-trained classifier and second by visually inspecting and critically discussing samples created using \tool. Our evaluation is limited to simple grayscale image classification datasets, where the rAAEs are easily trained. See \autoref{sec:threads} for a detailed discussion of the applicability of \tool.

\subsection{Quantitative Evaluation of \tool}
\label{sec:quantitative_evaluation}

We performed our experiments using four different DNN architectures as supervised models: 
A simple convolutional DNN~\cite{Chollet2020Mnistmodel}, a similar but fully connected DNN, 
a model consisting of Resnet-50~\cite{He2016Resnet} feature extraction and three fully connected layers for classification and 
lastly a Densenet-architecture~\cite{Huang2017Densenet}. 
Results are averaged over the four architectures, individual results are reported in the reproduction package.

We compare the predictions made for our ambiguous dataset  to the predictions made on nominal, non-ambiguous data, using the following metrics:

\noindent
\textbf{Top-1 / Regular Accuracy} Percentage of correctly classified inputs. 
We expect this to be considerably lower for ambiguous than for nominal samples, as choosing the correct (i.e., higher probability) class, even using an optimal model, is affected by chance.

\noindent
\textbf{Top-2 Accuracy} Percentage of inputs for which the true label is among the two classes with the highest predicted probability.
For samples which are truly ambiguous between two classes, we expect a well-trained model to achieve much better performance than on Top-1 accuracy (ideally, 100\%).

\noindent
\textbf{Top-Pair Accuracy} Novel metric for data known to be ambiguous between two classes, measured as the percentage of inputs for which the two most likely predicted classes equal the two true classes between which the input is ambiguous.
By definition Top-Pair accuracy is lower than or equal to Top-2 accuracy. It is an even stronger measure to show that the model is uncertain between exactly the two classes for which the true probabilistic label of the input shows nonzero probability. A specific example on how Top-Pair Accuracy is computed is provided in \autoref{appendix:top-pair}.

\noindent
\textbf{Entropy} Average entropy in the Softmax prediction arrays. Used as a metric to measure aleatoric uncertainty (and thus ambiguity) in related work~\cite{Mukhoti2021}.

In line with related literature~\cite{Mukhoti2021}, we focus our evaluations on models trained using a mixed-ambiguous dataset consisting of both nominal and ambiguous data.
This aims to make sure our ambiguous test sets are not OOD, and that thus the observed uncertainty primarily comes from the ambiguity in the data:
By adding a lot of data similar to the (ambiguous) test set to the training set, the vast majority of our ambiguous inputs is thus expected to be in-distribution, eradicating most of the epistemic uncertainty. The aleatoric uncertainty caused by the ambiguity of the data is however still there.
For completeness, we also run the evaluation on a model trained using only nominal data. With this model we expect even lower values of regular (top-1) accuracy on ambiguous data, as these are out-of-distribution, not just ambiguous.

\subsection{Quantitative Results}

\begin{table}
\centering
\begin{tabular}{@{}llcccc@{}}
\textbf{Training Set}            & \textbf{Test Set} & \textbf{Top-1 Acc} & \textbf{Top-2 Acc} & \textbf{Top-Pair Acc} & \textbf{Entropy} \\\midrule

\multicolumn{6}{c}{\textit{Our dataset for Fashion MNIST}}      \vspace{3px} \\

\multirow{2}{*}{mixed-ambiguous}   & ambiguous  & 0.51 & 0.94 & 0.87 & 1.33\\
                                         & nominal    & 0.88 & 0.96 & n.a. & 0.35\\
\multirow{2}{*}{clean}             & ambiguous  & 0.32 & 0.49 & 0.14 & 1.05\\
                                         & nominal    & 0.88 & 0.97 & n.a. & 0.33\\
    
\midrule
\multicolumn{6}{c}{\textit{Our dataset for MNIST}}      \vspace{3px} \\

\multirow{2}{*}{mixed-ambiguous}   & ambiguous  & 0.53 & 0.98 & 0.95 & 1.22\\
                                         & nominal    & 0.97 & 0.99 & n.a. & 0.15\\
\multirow{2}{*}{clean}             & ambiguous  & 0.42 & 0.64 & 0.32 & 0.73\\
                                         & nominal    & 0.97 & 0.99 & n.a. & 0.10\\
    
\midrule
\multicolumn{6}{c}{\textit{Baseline for mnist: AmbiguousMNIST by Mukhoti et. al.~\cite{Mukhoti2021}}}      \vspace{3px} \\

\multirow{2}{*}{mixed-ambiguous}   & ambiguous  & 0.72 & 0.91 & not calculable & 0.88\\
                                         & nominal    & 0.97 & 0.99 & n.a. & 0.12\\
\multirow{2}{*}{clean}             & ambiguous  & 0.65 & 0.85 & not calculable & 0.68\\
                                         & nominal    & 0.97 & 0.99 & n.a. & 0.11\\

\bottomrule
\end{tabular}
\caption{Evaluation of Ambiguity }
\label{tab:ambiguity_res}
\end{table}

The results of our experiments, averaged over all tested model architectures, are shown in \autoref{tab:ambiguity_res}. 
Results individually reported for each architecture are shown in \autoref{appendix:ambiguity-eval}.

We  noticed that the use of the mixed-ambiguous training sets reduces the model accuracy on nominal data only by a negligible amount: On MNIST, the corresponding accuracy is 96.98\% (97.42\% using a clean training set) and 88.43\% on FMNIST (88.37\% using a clean training set).
Thus, our ambiguous training datasets can be added to the nominal ones without hesitation.

Results indicate that our datasets are indeed suitable to induce ambiguity into the prediction process, as the generated data is  perceived as ambiguous by the DNN:
Top-1 accuracies for both case studies is around 50\%, but they increase almost to the levels of the nominal test set when considering Top-2 accuracies. 
Even Top-Pair accuracy, with values of 95.37\% and 86.71\% (on MNIST and FMNIST, respectively) are very high, showing that for the vast majority of test inputs, the two classes considered most likely by the well-trained DNN are exactly the classes between which we aimed to create ambiguity.
Consistently, entropy is substantially higher for ambiguous data than for nominal data.

Finally, we compared our ambiguous MNIST dataset against AmbiguousMNIST by Mukhoti et al.~\cite{Mukhoti2021}, the only publicly available dataset aiming to provide ambiguous data. Results are clearly in favour of our dataset.
Considering the models with mixed-ambiguous training sets\footnote{
When comparing against the baseline, we use the ambiguous training sets consistently with the test sets, i.e., the mixed-ambigous models used to assess our test sets also relied on our ambiguous training set, while the model used to assess the test set my Mukhoti et al. also was trained using their ambiguous training set.}\textsuperscript{,}\footnote{The clean-nominal case does not rely on, or test, any ambigous data. The values, which are reported for completeness, are thus expected to be identical for "our MNIST" and the baseline, with minimal differences being caused by randomness during training.}, our test dataset  has a lower Top-1 accuracy (53.31\% vs. 72.50\%), indicating that our dataset is harder (more ambiguous)
and has a higher Top-2 accuracy (97.99\% vs. 90.93\%) showing that our dataset contains more samples whose predicted class is amongst the 2 most likely labels.
Top-Pair accuracy cannot be computed for AmbiguousMNIST, as 37\% of its claimed “ambiguous” inputs have non-ambiguous labels.
Most strikingly, the average softmax entropy for AmbiguousMNIST is 0.88 (ours: 1.22), even though AmbiguousMNIST is created by actively selecting inputs with a high softmax entropy.

\subsection{Qualitative Discussion of \tool}
\def \unctypesubfigurewidth {0.19\linewidth}
\def \unctypeimgwidth {.8\linewidth}

\captionsetup{justification=centering}

\begin{figure}
\centering
\begin{subfigure}{\unctypesubfigurewidth}
  \centering
  \includegraphics[width=\unctypeimgwidth]{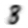}
  \caption{\textbf{good}\\(3/\underline{8})}
\end{subfigure}%
\begin{subfigure}{\unctypesubfigurewidth}
  \centering
  \includegraphics[width=\unctypeimgwidth]{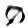}
  \caption{\textbf{good}\\(0/\underline{7})}
\end{subfigure}
\begin{subfigure}{\unctypesubfigurewidth}
  \centering
  \includegraphics[width=\unctypeimgwidth]{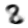}
  \caption{\textbf{good}\\(\underline{2}/8)}
\end{subfigure}%
\begin{subfigure}{\unctypesubfigurewidth}
  \centering
  \includegraphics[width=\unctypeimgwidth]{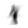}
  \caption{\textbf{ok}\\(\underline{1}/4)}
\end{subfigure}
\begin{subfigure}{\unctypesubfigurewidth}
  \centering
  \includegraphics[width=\unctypeimgwidth]{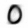}
  \caption{\textbf{bad}\\(\underline{0}/1)}
\end{subfigure}

\vspace{2mm}

\begin{subfigure}{\unctypesubfigurewidth}
  \centering
  \includegraphics[width=\unctypeimgwidth]{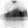}
  \caption{\textbf{good}\\(shirt/\\\underline{sneaker})}
\end{subfigure}
\begin{subfigure}{\unctypesubfigurewidth}
  \centering
  \includegraphics[width=\unctypeimgwidth]{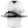}
  \caption{\textbf{good}\\(\underline{sandal}/\\shirt)}
\end{subfigure}
\begin{subfigure}{\unctypesubfigurewidth}
  \centering
  \includegraphics[width=\unctypeimgwidth]{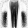}
  \caption{\textbf{good}\\(\underline{top}/\\trouser)}
\end{subfigure}
\begin{subfigure}{\unctypesubfigurewidth}
  \centering
  \includegraphics[width=\unctypeimgwidth]{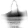}
  \caption{\textbf{good}\\(\underline{sneaker}/\\bag)}
\end{subfigure}
\begin{subfigure}{\unctypesubfigurewidth}
  \centering
  \includegraphics[width=\unctypeimgwidth]{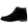}
  \caption{\textbf{bad}\\(\underline{sneaker}/\\ankle boot)}
\end{subfigure}

\captionsetup{justification=raggedright}
\caption{
Selected \emph{good} and \emph{bad} outputs of \tool, chosen to demonstrate strengths and weaknesses.
}
\label{fig:gen-amb}
\end{figure}

Some test samples generated using \tool, for both MNIST and FMNIST, are shown in \autoref{fig:gen-amb}.
They have been chosen to highlight different strengths and weaknesses that emerged during our qualitative manual review of 300 randomly selected images in our generated test sets per case study.

\paragraph{MNIST}  \tool (see Fig.~\ref{fig:gen-amb}a-e) is in general capable of combining features of different classes, where possible: ~\ref{fig:gen-amb}a and \ref{fig:gen-amb}c can both be seen as an 8, but the 8-shape was combined with a 3-shape or 2-shape, respectively.
For the combination between 0 and 7, shown in ~\ref{fig:gen-amb}b, only the upper (horizontal) part of the 7 was combined with the 0-shape, such that both a 7 and a 0 are clearly visible, making the class of the image ambiguous.
Fig. ~\ref{fig:gen-amb}d shows an edge case of an almost invalid image: Knowing that the image is supposed to be ambiguous between 1 and 4, one can identify both numbers. 
However, neither of them is clearly visible and the image may appear invalid to some humans.
Overall, we considered only few samples generated by \tool for MNIST as \emph{bad}, i.e., as clearly unambiguous or invalid.
An example of them is shown in Fig. ~\ref{fig:gen-amb}e. By most humans, this image would be recognized as an unambiguous 0. In fact, there's a barely visible, tilted line within the 0 which apparently was sufficient to trick the rAAEs discriminator into also assigning a high probability to digit 1. 

\paragraph{FMNIST} Realistic true ambiguity is not possible between most classes of FMNIST.
Hence, we assessed how well \tool performs at creating data that would trigger an ambiguous classification by humans, even though such data might be impossible to experience in the real world. Examples are given in Fig.~\ref{fig:gen-amb}f-j.
In most cases (e.g. Fig.~\ref{fig:gen-amb}f-h), the interpolations created by \tool show an overlay of two items of the two considered classes, with features combined only where possible.
We can also observe that some non-common features are removed, giving more weight to common features. For instance, in Fig. ~\ref{fig:gen-amb}i, the tip of the shoe, and the lower angles of the bag are barely noticeable, such that the image has indeed high similarity with both shoes and bags.
As a negative example we observe that, in some cases, it appears that  the overlay between the two considered classes is dominated by one one of them (such as Fig. ~\ref{fig:gen-amb}j, which would be seen as non-ambiguous ankle boot by most humans). 

\begin{tcolorbox}[colback=blue!5!white,colframe=blue!75!black,title=\textbf{Summary} (Evaluation of \tool-datasets)]
\tool successfully generated highly ambiguous data sets, with high prediction entropy, top-1 accuracy close to 50\% and top-2 accuracy close to 100\%, outperforming the ambiguous dataset previously produced by Mokhoti et al.~\cite{Mukhoti2021}.
\end{tcolorbox}

\section{Testing of Supervisors}
\label{benchmark}

We assess the capability of 16 supervisors\footnote{We are inclusive in our notion of supervisor: we consider also prioritization techniques that recognize unexpected inputs, as it is straightforward to adopt them to supervise a model.} to discriminate nominal from high-uncertainty inputs for  MNIST and FMNIST, each on 4 distinct test sets representing different root causes of mis-classifications, among which our ambiguous test set.

\subsection{Experimental Setup}
We performed our experiments using four different DNN architectures (explained in Section ~\ref{sec:quantitative_evaluation}) as supervised models.
Our training sets consist of both nominal and ambiguous data, to ensure that the ambiguous test data used later for testing is in-distribution. 
We then measure the capability of different supervisors to discriminate different types of high-uncertainty inputs from nominal data. 
We measure this using the \emph{area under the receiver operating characteristic curve} (AUC-ROC), a standard, threshold-independent metric.

We assess the supervisors using the following test sets:
\emph{Invalid} test sets, where we use MNIST images as inputs to models trained for FMNIST and vice-versa,
\emph{corrupted} test sets available from related work (MNIST-C~\cite{Mu2019} and FMNIST-c~\cite{weiss2022simpleTip}),
\emph{adversarial} data, created using 4 different attacks~\cite{madry2017adversarialPGD, Kurakin2018adversarialBIM, moosavi2016adversarialDeepfool, Goodfellow2014explainingAdvFGSM} and lastly the \emph{ambiguous} test sets generated by \tool.
Adversarial test sets were not used with ensembles, as an ensemble does not rely on the (single) model targeted by the considered adversarial test generation techniques.

To account for random influences during training, such as initial model weights, we ran the experiments for each DNN architecture 5 times. 
Results reported are the means of the observed results. 

\subsection{Tested Supervisors}
To avoid unnecessary redundancy, our description of the tested supervisors is brief and we refer to the corresponding papers for a detailed presentation.
Our terminology, implementation and configuration of  the first three supervisors described below, i.e., Softmax, MC-Dropout and Ensembles, are based on the material released in our recent empirical studies~\cite{Weiss2021, Weiss2022Stvr}.

\noindent
\textbf{Plain Softmax}
Based solely on the softmax output array of a DNN prediction, these approaches provide very fast and easy to compute supervision: \emph{Max. Softmax}, highest softmax value as confidence~\cite{Hendrycks2016},  \emph{Prediction-Confidence Score (PCS)}, the difference between the two highest softmax values~\cite{Zhang2020}, \emph{DeepGini}, the complement of the softmax vector squared norm~\cite{feng2020deepgini}, and finally the \emph{entropy} of the values in the predicted softmax probabilities~\cite{Weiss2021Uwiz}.

\noindent
\textbf{Monte-Carlo Dropout (MC-Dropout)}~\cite{Gal2016, Gal2016a}
Enabling the randomness of dropout layers at prediction time, and sampling multiple  randomized samples allows the inference of an output distribution, hence of an uncertainty quantification. 
We use the quantifiers \emph{Variation Ratio (VR)}, \emph{Mutual Information (MI)},
\emph{Predictive Entropy} (PI), 
or simply the highest value of the mean of the predicted softmax likelihoods (\emph{Mean-Softmax, MS}).

\noindent
\textbf{Ensembles}~\cite{Lakshminarayanan2017} Similar to MC-Dropout, uncertainty is inferred from samples, but randomness is induced by training multiple models  (under random influences such as initial weights) and collecting predictions from all of them. Here, we use the quantifiers MI, PI and MS, on an Ensemble consisting of 20 models.

\noindent
\textbf{Dissector}~\cite{Wang2020Dissector}
On a trained model, for each layer, a \emph{submodel} (more specifically, a perceptron) is trained, predicting the label directly from the activations of the given layer.
From these outputs, the \emph{support value} for each of the submodels for the prediction made by the final layer is calculated, and the overall \emph{prediction validity} value is calculated as a weighted average of the per-layer support values. 

\noindent
\textbf{Autoencoders} AEs can be used as OOD detectors: If the reconstruction error of a well-trained AE for a given input is high, it is likely not to be sufficiently represented in the training data. Stocco et. al.,~\cite{Stocco2020} proposed to use such OOD detection technique as DNN supervisor. Based on their findings, we use a \emph{variational autoencoder}~\cite{Kingma2013}.

\noindent
\textbf{Surprise Adequacy} This approach detects inputs that are \emph{surprising}, i.e., for which the observed DNN activation pattern is OOD w.r.t. the ones observed on the training data.
We consider three techniques to quantify surprise adequacy: 
\emph{LSA}~\cite{Kim2018}, where surprise is calculated based on a kernel-density estimator fitted on the training activations of the predicted class, \emph{MDSA}~\cite{Kim2020MahalanobisSA}, where surprise is calculated based on the Mahalanobis distance between the tested input's activations and the training activations of the predicted class, and \emph{DSA}~\cite{Kim2018} which is calculated as the ratio between two Euclidean distances: the distance between the tested input and the closest training set activation in the predicted class, and the distance between the latter activation and the closest training set activation from another class.
As DSA is computationally intensive, growing linearly in the number of training samples, we follow a recent proposal to consider only 30\% of the training data~\cite{Weiss2021-SA}. 

Our comparison includes most of the popular supervisors used in recent software engineering literature.
Some of the excluded techniques do not provide a single, continuous uncertainty score and no AUC-ROC can thus be calculated for them~\cite{Catak2021MultiScore, Mukhoti2021, postels2020quantifying }, %
or they are not applicable to the image classification domain~\cite{Hussain2022}.
With its 16 tested supervisors, two case studies and four different data-centric root causes of DNN faults, 
our study is -- to the best of our knowledge -- by far the most extensive of its kind.

\subsection{Results}

    \newcolumntype{Y}{>{\centering\arraybackslash}X}
    
    \begin{table*}[t]
    \begin{tabularx}{\linewidth}{l|YYYY|YYYY}
\toprule
{} & \multicolumn{4}{c}{mnist} & \multicolumn{4}{c}{fmnist} \\
{} & amb. & adv. & corr. & inv. & amb. & adv. & corr. & inv. \\
\midrule

\multicolumn{9}{l}{\small \textit{Plain Softmax Supervisors}}\vspace{2px}\\Max. Softmax       &      0.96 &        0.79 &      0.78 &    0.79 &      0.91 &        0.61 &      0.71 &    0.73 \\
PCS                &      0.96 &        0.79 &      0.78 &    0.79 &      0.91 &        0.61 &      0.70 &    0.72 \\
Softmax Entropy    &      0.97 &        0.79 &      0.78 &    0.79 &      0.92 &        0.61 &      0.72 &    0.74 \\
DeepGini           &      0.96 &        0.79 &      0.78 &    0.79 &      0.92 &        0.61 &      0.71 &    0.73 \vspace{2px}\\
\multicolumn{9}{l}{\small \textit{Monte-Carlo Dropout Supervisors (Softmax-based, except for VR)}}\vspace{2px}\\MC-Dropout (VR)    &      0.79 &        0.69 &      0.65 &    0.72 &      0.76 &        0.62 &      0.66 &    0.72 \\
MC-Dropout (MS)    &      0.96 &        0.79 &      0.80 &    0.80 &      0.91 &        0.61 &      0.73 &    0.77 \\
MC-Dropout (MI)    &      0.87 &        0.78 &      0.81 &    0.83 &      0.73 &        0.61 &      0.78 &    0.86 \\
MC-Dropout (PE)    &      0.96 &        0.79 &      0.80 &    0.80 &      0.91 &        0.61 &      0.74 &    0.79 \vspace{2px}\\
\multicolumn{9}{l}{\small \textit{Deep Ensemble Supervisors (Softmax-based)}}\vspace{2px}\\Deep Ensemble (MS) &      0.97 &        n.a. &      0.84 &    0.85 &      0.90 &        n.a. &      0.75 &    0.64 \\
Deep Ensemble (MI) &      0.84 &        n.a. &      0.84 &    0.88 &      0.57 &        n.a. &      0.76 &    0.70 \\
Deep Ensemble (PE) &      0.97 &        n.a. &      0.83 &    0.84 &      0.89 &        n.a. &      0.77 &    0.66 \vspace{2px}\\
\multicolumn{9}{l}{\small \textit{Other Supervisors}}\vspace{2px}\\Dissector          &      0.95 &        0.79 &      0.76 &    0.79 &      0.88 &        0.68 &      0.72 &    0.75 \\
DSA                &      0.48 &        0.93 &      0.87 &    0.98 &      0.31 &        0.85 &      0.85 &    0.90 \\
LSA                &      0.17 &        0.78 &      0.73 &    0.77 &      0.16 &        0.75 &      0.74 &    0.86 \\
MDSA               &      0.31 &        0.94 &      0.87 &    0.98 &      0.32 &        0.86 &      0.83 &    0.95 \\
Autoencoder        &      0.62 &        0.95 &      0.84 &    1.00 &      0.53 &        0.80 &      0.77 &    0.49 \\
\bottomrule
\end{tabularx}
\caption{Supervisors performance at discriminating nominal from high-uncertainty inputs (AUC-ROC), averaged over all architectures.}
\label{tab:mean_auc_roc}
\end{table*}
Overall observed results (averaged over all models) are are presented in \autoref{tab:mean_auc_roc}. Per-Architecture results with the corresponding standard deviations are shown in \autoref{appendix:supervisor-eval}.

\paragraph{Ambiguous Data} We can observe that the predicted softmax likelihoods capture aleatoric uncertainty pretty well.
Thus, not only do Max. Softmax, DeepGini, PCS, Softmax Entropy perform well at discriminating ambiguous from nominal data, but also supervisors that rely on the softmax predictions indirectly, such as Dissector, or the MS, MI and PE quantifiers on samples collected using MC-Dropout or DeepEnsembles.
DSA, LSA, MDSA and Autoencoders are not capable of detecting ambiguity, and barely any of their AUC-ROCs exceeds the 0.5 value expected from a random classifier on a balanced dataset.
MDSA, LSA and DSA show particularly low values, which confirms that  they do only one job -- detecting OOD, not ambiguous data -- but they do it well (in our experimental design, ambiguous data is in-distribution by construction, while adversarial, corrupted and invalid data is OOD).

\paragraph{Adversarial Data} 
The surprise adequacy based supervisors and the autoencoder reliably detected the unknown patterns in the input, discriminating adversarial from nominal data.
Softmax-based supervisors showed good results on MNIST, but less so on FMNIST.
Cleary, the adversarial sample detection capabilities of Softmax-based supervisors depend critically on the choice of adversarial data: With minimal perturbations, just strong enough to trigger a misclassification, softmax-based metrics can easily detect them, as the maximum of the predicted softmax likelihood is artificially reduced by the adversarial technique being used. However, one could apply stronger attacks, increasing the predicted likelihood of the wrong class close to 100\%, which would make Softmax-based supervisors ineffective.
Specific attacks against the other supervisors, i.e., the OOD detection based approaches (surprise adequacies and autoencoders) and Dissector, might also be possible in theory, but they are clearly much harder.

\paragraph{Corrupted Data} Most approaches perform comparably well, with the exception of DSA on FMNIST, which shows  superior performance, with an average AUC-ROC value more than $.1$ higher than most other supervisors.
DNNs are known to sometimes map OOD data points close to feature representations of in-distribution points (known as \emph{feature collapse})~\cite{VanAmersfoort2021featureCollapse}, 
thus leading to softmax output distributions similar to the ones of in-distribution images.
This impacts negatively the OOD detection capability of Softmax-based supervisors (such as Max. Softmax, MC-Dropout, Ensembles or Dissector), especially in cases with a feature-rich training set, such as FMNIST.

\paragraph{Invalid Data}

The best result in invalidity detection was achieved using the autoencoder's reconstruction error, which identified FMNIST inputs given to an MNIST classifier with an AUC-ROC of $\approx1.00$ (thus with almost perfect accuracy). Clearly, reconstructions of images with higher feature complexity than the ones represented in the training set consistently lead to high reconstruction errors and thus provided a very reliable outlier detection. The autoencoder was however incapable of detecting MNIST images given to an FMNIST classifier (AUC-ROC of $\approx0.49$, similar to a random classifier). Here, it seems that an autoencoder trained on a high feature-complexity  training set would also learn to reconstruct low feature-complexity  inputs accurately.
DSA and MDSA, showed a similar effect, also providing clearly inferior results in the FMNIST case study compared to MNIST, although here the drop in performance was less dramatic. Also, similar to corrupted data, most likely due to feature collapse, the performance of supervisors relying on softmax likelihoods suffers dramatically.

\paragraph{Discussion}
Related literature suggests that no single supervisor performs well under all conditions~\cite{Zhang2018, Weiss2021, Weiss2022Stvr, Catak2021MultiScore}, 
and some works even suggest that certain supervisors are not capable to detect anything but aleatoric uncertainty (e.g.  Softmax Entropy~\cite{Mukhoti2021} or MC-Dropout~\cite{osband2016risk}).
Our evaluation, to the best of our knowledge, is the first one which compares supervisors on four different uncertainty-inducing test sets.
We found that softmax-based approaches (including MC-Dropout, Ensembles and Dissector) are \emph{effective} on all four types of test sets, i.e., their detection capabilities reliably exceed the performance expected from a random classifier. 
They do have their primary strength in the detection of ambiguous data, where the other, OOD focused techniques are naturally ineffective,
but they are actually an inferior choice when targeting epistemic uncertainty.
To detect corrupted inputs, DSA exhibited the best performance, but due to its high computational complexity it may not be suitable to all domains. The much faster MDSA may offer a good trade-off between detection capability and runtime complexity.
Regarding invalid inputs, on low-feature problems, where invalid samples are expected to be more complex than nominal inputs, AEs provide a fast approach with the additional advantage that it does not rely on the supervised model directly, but only on its training set, which may facilitate maintenance and continuous development. 
For problems where the nominal inputs are rich of diverse features, an AE is not a valid option. However,  our results again suggest MDSA as a reliable and fast alternative supervisor.
For what regards inputs created by adversarial attacks, softmax-based approaches are easily deceived, being hence of limited practical utility. 
On the other hand, OOD detectors, such as surprise adequacy metrics and AEs, or Dissector can provide a more reliable detection performance against standard adversarial attacks. 
Of course,  these supervisors are not immune from particularly malicious attackers that target them specifically. Here, the reader can refer to the wide range of research discussing defenses against adversarial attacks (survey provided by Akhtar et. al.~\cite{akhtar2021advancesAdversarial}).

    \paragraph{Stability of Results}
    We found that our results are barely sensitive to random influences due to training: 
    Out of 488 reported mean AUC-ROCs (4 architectures, 8 test sets, 16 MPs, averaged over 5 runs\footnote{Results reported in the appendix (tables ~\ref{tab:auc_roc_simplecnn}, ~\ref{tab:auc_roc_fullyconnectednet}, ~\ref{tab:auc_roc_densenet} and ~\ref{tab:auc_roc_resnet}) })
    most of them showed a negligible standard deviation:
    The average observed standard deviation was 0.015, the highest one was 0.124, 
    only 114 were larger than 0.02, 
    only 29 were larger than 0.05, all of which correspond to results with low mean AUC-ROC (\textless 0.9). 
    The latter differences do not influence the overall observed tendencies.

\begin{tcolorbox}[colback=blue!5!white,colframe=blue!75!black,title=\textbf{Summary} (Comparison of Misclassification Detectors)]
We assessed 16 supervisors on their capability to discriminate between nominal inputs and inputs which are ambiguous, adversarial, corrupted or invalid. 
For every category, we identified supervisors which perform particularly well, but we also found that to target all types of high-uncertainty inputs developers of DLS will have to rely on multiple, diverse supervisors.
\end{tcolorbox}

\section{Threats to validity}
\label{sec:threads}

\textbf{External validity}: We conducted our study on misclassification predition considering two standard case studies, MNIST and Fashion-MNIST. While our observations may not generalize to more challenging, high uncertainty datasets, the choice of two simple datasets with easily understandable features, allowed us to achieve a clear and sharp separation of the reasons for failures, which may not be the case when dealing with more complex datasets. On the other hand, we recognize the importance of replicating and extending this study considering additional datasets. To support such replications we provide all our experimental material as open source/data.

\textbf{Internal validity}: The supervisors being compared include hyper-parameters that require some tuning. Whenever possible, we reused the original values and followed the guidelines proposed by the authors of the considered approaches. We also conducted a few preliminary experiments to validate and fine tune such hyper-parameters. However, the configurations used in our experiment could be suboptimal for some supervisor.

\textbf{Conclusion validity}: We repeated our experiments 5 times to mitigate the non determinism associated with the DNN training process. While this might look like a low number of repetitions, we  checked the standard deviation across such repetitions and found that it was negligible or small in all cases. To amount for the influence of the DNN architecture, we performed our experiments on 4 completely different DNN architectures, obtaining overall consistent findings.

\section{Conclusion}
\label{conclusion}

This paper brings two major advances to the field of DNN supervision testing:
First, we proposed \tool, a novel technique to create labeled ambiguous images in a way that is independent of the tested model and of its supervisor, and we generated pre-compiled ambiguous datasets for two of the most popular case studies in DNN testing research, MNIST and Fashion-MNIST.

Using four different metrics, we were able to verify the validity and ambiguity of our datasets, and we further investigated \textit{how} \tool achieves ambiguity based on a qualitative analysis.
On the four considered quantitative indicators, \tool clearly outperformed AmbiguousMNIST, the only similar-purposed dataset in the literature.

We assessed the capabilities of 16 DNN supervisors at discriminating nominal from ambiguous, adversarial, corrupted and invalid inputs.
To the best of our knowledge, this is  not only the largest empirical case study comparing DNN supervisors in the literature, it is also the first one to do so by specifically targeting \emph{four} distinct and clearly separable data-centric root causes of DNN faults.
Our results show that softmax-based approaches (including MC-Dropout and Ensembles) work very well at detecting ambiguity, but have clear disadvantages when it comes to adversarial, corrupted, and invalid inputs. 
OOD detection techniques, such as surprise adequacy or autoencoder-based supervisors, often provide a better detection performance with the targeted types of high-uncertainty inputs. However, these approaches are incapable of detecting in-distribution ambiguous inputs.

DNN developers can use the ambiguous datasets created by \tool to  assess novel DNN supervisors on their capability to detect aleatoric uncertainty. They can also use our tool to evaluate test prioritization approaches on their capability to prioritize ambiguous inputs (depending on the developers' objectives, high priority  is desired to identify inputs that are likely to be misclassified during testing; low priority is desired to exclude inputs with probabilistic labels from the training set).

As future work, we plan to investigate the concept of true ambiguity for regression problems. This is relevant in domains, such as self-driving cars and robotics, where the DNN output is a continuous signal for an actuator.
This problem is particularly appealing as all the approaches in our study that worked well at detecting ambiguity are based on softmax and thus are not applicable to regression problems. 

Additionally, a comprehensive human experiment evaluating and comparing the ambiguity of data in no\-mi\-nal datasets, data created using \tool, and data generated by other approaches would help to better understand the nature of these datasets.

\section{Data Availability}
During the review\footnote{Upon acceptance of this paper, we will release the code and generated artifacts using a range of common formats and through different platforms (such as Huggingface-datasets~\cite{Lhoest2021HuggingfaceDatasets}, Github and Zenodo), using permissive licenses.}, the artifacts are made available to the reviewers using the following link:

\[\text{\url{https://mweiss.ch/emse-ambiguity-artifact}}\]

\section*{Conflict of interest}
The authors declare that they have no conflict of interest.

\bibliographystyle{spmpsci}      %
\bibliography{main}   %

\begin{thebibliography}{10}
\providecommand{\url}[1]{{#1}}
\providecommand{\urlprefix}{URL }
\expandafter\ifx\csname urlstyle\endcsname\relax
  \providecommand{\doi}[1]{DOI~\discretionary{}{}{}#1}\else
  \providecommand{\doi}{DOI~\discretionary{}{}{}\begingroup
  \urlstyle{rm}\Url}\fi

\bibitem{akhtar2021advancesAdversarial}
Akhtar, N., Mian, A., Kardan, N., Shah, M.: Advances in adversarial attacks and
  defenses in computer vision: A survey.
\newblock IEEE Access \textbf{9}, 155161--155196 (2021)

\bibitem{VanAmersfoort2021featureCollapse}
van Amersfoort, J., Smith, L., Jesson, A., Key, O., Gal, Y.: On feature
  collapse and deep kernel learning for single forward pass uncertainty.
\newblock arXiv preprint arXiv:2102.11409  (2021)

\bibitem{Aroyo2021}
Aroyo, L., Paritoshs, P.: Uncovering unknown unknowns in machine learning
  (2021).
\newblock
  \urlprefix\url{https://ai.googleblog.com/2021/02/uncovering-unknown-unknowns-in-machine.html}

\bibitem{Ayhan2018}
Ayhan, M.S., Berens, P.: Test-time data augmentation for estimation of
  heteroscedastic aleatoric uncertainty in deep neural networks.
\newblock Presented at "Medical Imaging with Deep Learning 2018", Amsterdam
  (2018).
\newblock Available on OpenReview

\bibitem{Berend2020}
Berend, D., Xie, X., Ma, L., Zhou, L., Liu, Y., Xu, C., Zhao, J.: Cats are not
  fish: Deep learning testing calls for out-of-distribution awareness.
\newblock In: The 35th IEEE/ACM International Conference on Automated Software
  Engineering. Association for Computing Machinery, New York, NY, USA (2020)

\bibitem{bjarnadottir2019climate}
Bjarnadottir, S., Li, Y., Stewart, M.G.: Climate adaptation for housing in
  hurricane regions.
\newblock In: Climate Adaptation Engineering, pp. 271--299. Elsevier (2019)

\bibitem{brown2021deep}
Brown, J.M., Leontidis, G.: Deep learning for computer-aided diagnosis in
  ophthalmology: a review.
\newblock State of the Art in Neural Networks and their Applications pp.
  219--237 (2021)

\bibitem{Byun2020}
Byun, T., Rayadurgam, S.: Manifold for machine learning assurance.
\newblock In: Proceedings of the ACM/IEEE 42nd International Conference on
  Software Engineering: New Ideas and Emerging Results, pp. 97--100 (2020)

\bibitem{Catak2021}
Catak, F.O., Yue, T., Ali, S.: Prediction surface uncertainty quantification in
  object detection models for autonomous driving  (2021)

\bibitem{Catak2021MultiScore}
Catak, F.O., Yue, T., Ali, S.: Uncertainty-aware prediction validator in deep
  learning models for cyber-physical system data.
\newblock ACM Transactions on Software Engineering and Methodology  (2021)

\bibitem{Chollet2020Mnistmodel}
Chollet, F.: Keras documentation: Simple mnist convnet (2020).
\newblock \urlprefix\url{https://keras.io/examples/vision/mnist_convnet/}

\bibitem{Clements2019}
Clements, W.R., Delft, B.V., Robaglia, B.M., Slaoui, R.B., Toth, S.: Estimating
  risk and uncertainty in deep reinforcement learning  (2019)

\bibitem{davidson2021malaria}
Davidson, M.S., Andradi-Brown, C., Yahiya, S., Chmielewski, J., O’Donnell,
  A.J., Gurung, P., Jeninga, M.D., Prommana, P., Andrew, D.W., Petter, M.,
  et~al.: Automated detection and staging of malaria parasites from cytological
  smears using convolutional neural networks.
\newblock Biological imaging \textbf{1} (2021)

\bibitem{Dola2021}
Dola, S., Dwyer, M.B., Soffa, M.L.: Distribution-aware testing of neural
  networks using generative models pp. 226--237 (2021)

\bibitem{dunn2021exposing}
Dunn, I., Pouget, H., Kroening, D., Melham, T.: Exposing previously
  undetectable faults in deep neural networks.
\newblock In: Proceedings of the 30th ACM SIGSOFT International Symposium on
  Software Testing and Analysis, pp. 56--66 (2021)

\bibitem{feng2020deepgini}
Feng, Y., Shi, Q., Gao, X., Wan, J., Fang, C., Chen, Z.: Deepgini: prioritizing
  massive tests to enhance the robustness of deep neural networks.
\newblock In: Proceedings of the 29th ACM SIGSOFT International Symposium on
  Software Testing and Analysis, pp. 177--188 (2020)

\bibitem{Gal2016a}
Gal, Y.: Uncertainty in deep learning.
\newblock Ph.D. thesis, University of Cambridge (2016)

\bibitem{Gal2016}
Gal, Y., Ghahramani, Z.: Dropout as a bayesian approximation: Representing
  model uncertainty in deep learning.
\newblock In: Proceedings of the 33rd International Conference on International
  Conference on Machine Learning - Volume 48, ICML'16, pp. 1050--1059. JMLR.org
  (2016).
\newblock \urlprefix\url{http://dl.acm.org/citation.cfm?id=3045390.3045502}

\bibitem{Goodfellow2014}
Goodfellow, I., Pouget-Abadie, J., Mirza, M., Xu, B., Warde-Farley, D., Ozair,
  S., Courville, A., Bengio, Y.: Generative adversarial nets.
\newblock Advances in neural information processing systems \textbf{27} (2014)

\bibitem{Goodfellow2014explainingAdvFGSM}
Goodfellow, I.J., Shlens, J., Szegedy, C.: Explaining and harnessing
  adversarial examples.
\newblock arXiv preprint arXiv:1412.6572  (2014)

\bibitem{He2016Resnet}
He, K., Zhang, X., Ren, S., Sun, J.: Deep residual learning for image
  recognition.
\newblock In: Proceedings of the IEEE conference on computer vision and pattern
  recognition, pp. 770--778 (2016)

\bibitem{hell2021monitoring}
Hell, F., Hinz, G., Liu, F., Goyal, S., Pei, K., Lytvynenko, T., Knoll, A.,
  Yiqiang, C.: Monitoring perception reliability in autonomous driving:
  Distributional shift detection for estimating the impact of input data on
  prediction accuracy.
\newblock In: Computer Science in Cars Symposium, pp. 1--9 (2021)

\bibitem{Hendrycks2018}
Hendrycks, D., Dietterich, T.: Benchmarking neural network robustness to common
  corruptions and perturbations.
\newblock International Conference on Learning Representations  (2018)

\bibitem{Hendrycks2016}
Hendrycks, D., Gimpel, K.: A baseline for detecting misclassified and
  out-of-distribution examples in neural networks  (2016)

\bibitem{Henriksson2019}
Henriksson, J., Berger, C., Borg, M., Tornberg, L., Englund, C., Sathyamoorthy,
  S.R., Ursing, S.: Towards structured evaluation of deep neural network
  supervisors.
\newblock In: 2019 {IEEE} International Conference On Artificial Intelligence
  Testing ({AITest}). {IEEE} (2019).
\newblock \doi{10.1109/aitest.2019.00-12}

\bibitem{Henriksson2019a}
Henriksson, J., Berger, C., Borg, M., Tornberg, L., Sathyamoorthy, S.R.,
  Englund, C.: Performance analysis of out-of-distribution detection on various
  trained neural networks.
\newblock In: 2019 45th Euromicro Conference on Software Engineering and
  Advanced Applications (SEAA), pp. 113--120. IEEE (2019)

\bibitem{DeHond2022HealthcareAiGuidelines}
de~Hond, A.A., Leeuwenberg, A.M., Hooft, L., Kant, I.M., Nijman, S.W., van Os,
  H.J., Aardoom, J.J., Debray, T., Schuit, E., van Smeden, M., et~al.:
  Guidelines and quality criteria for artificial intelligence-based prediction
  models in healthcare: a scoping review.
\newblock npj Digital Medicine \textbf{5}(1), 1--13 (2022)

\bibitem{Huang2017Densenet}
Huang, G., Liu, Z., Van Der~Maaten, L., Weinberger, K.Q.: Densely connected
  convolutional networks.
\newblock In: Proceedings of the IEEE conference on computer vision and pattern
  recognition, pp. 4700--4708 (2017)

\bibitem{Humbatova2020}
Humbatova, N., Jahangirova, G., Bavota, G., Riccio, V., Stocco, A., Tonella,
  P.: Taxonomy of real faults in deep learning systems.
\newblock In: Proceedings of the ACM/IEEE 42nd International Conference on
  Software Engineering, pp. 1110--1121 (2020)

\bibitem{Hussain2022}
Hussain, M., Ali, N., Hong, J.E.: Deepguard: a framework for safeguarding
  autonomous driving systems from inconsistent behaviour.
\newblock Automated Software Engineering \textbf{29}(1), 1--32 (2022)

\bibitem{Kang2020}
Kang, S., Feldt, R., Yoo, S.: Sinvad: Search-based image space navigation for
  dnn image classifier test input generation.
\newblock In: Proceedings of the IEEE/ACM 42nd International Conference on
  Software Engineering Workshops, pp. 521--528 (2020)

\bibitem{Karimi2019}
Karimi, H., Derr, T., Tang, J.: Characterizing the decision boundary of deep
  neural networks  (2019)

\bibitem{Kim2018}
Kim, J., Feldt, R., Yoo, S.: Guiding deep learning system testing using
  surprise adequacy  (2018)

\bibitem{Kim2020MahalanobisSA}
Kim, J., Ju, J., Feldt, R., Yoo, S.: Reducing dnn labelling cost using surprise
  adequacy: An industrial case study for autonomous driving.
\newblock In: Proceedings of the 28th ACM Joint Meeting on European Software
  Engineering Conference and Symposium on the Foundations of Software
  Engineering, pp. 1466--1476 (2020)

\bibitem{Kim2021}
Kim, S., Yoo, S.: Multimodal surprise adequacy analysis of inputs for natural
  language processing dnn models.
\newblock In: 2021 2021 IEEE/ACM International Conference on Automation of
  Software Test (AST) (AST), pp. 80--89. IEEE Computer Society, Los Alamitos,
  CA, USA (2021).
\newblock \doi{10.1109/AST52587.2021.00017}.
\newblock
  \urlprefix\url{https://doi.ieeecomputersociety.org/10.1109/AST52587.2021.00017}

\bibitem{Kingma2013}
Kingma, D.P., Welling, M.: Auto-encoding variational bayes.
\newblock arXiv preprint arXiv:1312.6114  (2013)

\bibitem{Kurakin2018adversarialBIM}
Kurakin, A., Goodfellow, I.J., Bengio, S.: Adversarial examples in the physical
  world.
\newblock In: Artificial intelligence safety and security, pp. 99--112. Chapman
  and Hall/CRC (2018)

\bibitem{Lakshminarayanan2017}
Lakshminarayanan, B., Pritzel, A., Blundell, C.: Simple and scalable predictive
  uncertainty estimation using deep ensembles.
\newblock In: Advances in neural information processing systems, pp. 6402--6413
  (2017)

\bibitem{LeCun1998}
LeCun, Y., Bottou, L., Bengio, Y., Haffner, P.: Gradient-based learning applied
  to document recognition.
\newblock Proceedings of the IEEE \textbf{86}(11), 2278--2324 (1998)

\bibitem{Lhoest2021HuggingfaceDatasets}
Lhoest, Q., Villanova~del Moral, A., Jernite, Y., Thakur, A., von Platen, P.,
  Patil, S., Chaumond, J., Drame, M., Plu, J., Tunstall, L., Davison, J.,
  {\v{S}}a{\v{s}}ko, M., Chhablani, G., Malik, B., Brandeis, S., Le~Scao, T.,
  Sanh, V., Xu, C., Patry, N., McMillan-Major, A., Schmid, P., Gugger, S.,
  Delangue, C., Matussi{\`e}re, T., Debut, L., Bekman, S., Cistac, P.,
  Goehringer, T., Mustar, V., Lagunas, F., Rush, A., Wolf, T.: Datasets: A
  community library for natural language processing.
\newblock In: Proceedings of the 2021 Conference on Empirical Methods in
  Natural Language Processing: System Demonstrations, pp. 175--184. Association
  for Computational Linguistics, Online and Punta Cana, Dominican Republic
  (2021).
\newblock \urlprefix\url{https://aclanthology.org/2021.emnlp-demo.21}

\bibitem{Lines2019Disentangling}
Lines, D.: Disentangling sources of uncertainty for active exploration.
\newblock Master's thesis, Department of Engineering, University of Cambridge
  (2019)

\bibitem{madry2017adversarialPGD}
Madry, A., Makelov, A., Schmidt, L., Tsipras, D., Vladu, A.: Towards deep
  learning models resistant to adversarial attacks.
\newblock arXiv preprint arXiv:1706.06083  (2017)

\bibitem{Makhzani2015}
Makhzani, A., Shlens, J., Jaitly, N., Goodfellow, I., Frey, B.: Adversarial
  autoencoders.
\newblock arXiv preprint arXiv:1511.05644  (2015)

\bibitem{moosavi2016adversarialDeepfool}
Moosavi-Dezfooli, S.M., Fawzi, A., Frossard, P.: Deepfool: a simple and
  accurate method to fool deep neural networks.
\newblock In: Proceedings of the IEEE conference on computer vision and pattern
  recognition, pp. 2574--2582 (2016)

\bibitem{Mu2019}
Mu, N., Gilmer, J.: Mnist-c: A robustness benchmark for computer vision.
\newblock CoRR  (2019)

\bibitem{Mukhoti2021}
Mukhoti, J., Kirsch, A., van Amersfoort, J., Torr, P.H.S., Gal, Y.:
  Deterministic neural networks with appropriate inductive biases capture
  epistemic and aleatoric uncertainty  (2021).
\newblock Presented at the ICML UDL 2021 Workshop (non-archival)

\bibitem{Odena2019}
Odena, A., Olsson, C., Andersen, D., Goodfellow, I.: {T}ensor{F}uzz: Debugging
  neural networks with coverage-guided fuzzing.
\newblock In: K.~Chaudhuri, R.~Salakhutdinov (eds.) Proceedings of the 36th
  International Conference on Machine Learning, \emph{Proceedings of Machine
  Learning Research}, vol.~97, pp. 4901--4911. PMLR, Long Beach, California,
  USA (2019).
\newblock \urlprefix\url{http://proceedings.mlr.press/v97/odena19a.html}

\bibitem{osband2016risk}
Osband, I.: Risk versus uncertainty in deep learning: Bayes, bootstrap and the
  dangers of dropout.
\newblock In: NIPS workshop on bayesian deep learning, vol. 192 (2016)

\bibitem{postels2020quantifying}
Postels, J., Blum, H., Cadena, C., Siegwart, R., Van~Gool, L., Tombari, F.:
  Quantifying aleatoric and epistemic uncertainty using density estimation in
  latent space.
\newblock arXiv preprint arXiv:2012.03082  (2020)

\bibitem{Rauber2017foolbox}
Rauber, J., Brendel, W., Bethge, M.: Foolbox: A python toolbox to benchmark the
  robustness of machine learning models.
\newblock In: Reliable Machine Learning in the Wild Workshop, 34th
  International Conference on Machine Learning (2017).
\newblock \urlprefix\url{http://arxiv.org/abs/1707.04131}

\bibitem{Riccio2020}
Riccio, V., Jahangirova, G., Stocco, A., Humbatova, N., Weiss, M., Tonella, P.:
  Testing machine learning based systems: a systematic mapping.
\newblock Empirical Software Engineering  (2020)

\bibitem{Riccio2020Janus}
Riccio, V., Tonella, P.: Model-based exploration of the frontier of behaviours
  for deep learning system testing.
\newblock In: Proceedings of the 28th ACM Joint Meeting on European Software
  Engineering Conference and Symposium on the Foundations of Software
  Engineering, pp. 876--888 (2020)

\bibitem{Samek2017}
Samek, W., Wiegand, T., M{\"u}ller, K.R.: Explainable artificial intelligence:
  Understanding, visualizing and interpreting deep learning models.
\newblock arXiv preprint arXiv:1708.08296  (2017)

\bibitem{Seca2021}
Seca, D.: A review on oracle issues in machine learning.
\newblock arXiv preprint arXiv:2105.01407  (2021)

\bibitem{Stocco2020}
Stocco, A., Weiss, M., Calzana, M., Tonella, P.: Misbehaviour prediction for
  autonomous driving systems.
\newblock In: Proceedings of 42nd International Conference on Software
  Engineering, p. 12 pages. ACM (2020)

\bibitem{Tian2018Deeptest}
Tian, Y., Pei, K., Jana, S., Ray, B.: Deeptest: Automated testing of
  deep-neural-network-driven autonomous cars.
\newblock In: Proceedings of the 40th international conference on software
  engineering, pp. 303--314 (2018)

\bibitem{trappenberg2000classification}
Trappenberg, T.P., Back, A.D.: A classification scheme for applications with
  ambiguous data.
\newblock In: Proceedings of the IEEE-INNS-ENNS International Joint Conference
  on Neural Networks. IJCNN 2000. Neural Computing: New Challenges and
  Perspectives for the New Millennium, vol.~6, pp. 296--301. IEEE (2000)

\bibitem{Wang2020Dissector}
Wang, H., Xu, J., Xu, C., Ma, X., Lu, J.: Dissector: Input validation for deep
  learning applications by crossing-layer dissection.
\newblock In: Proceedings of 42nd International Conference on Software
  Engineering. ACM (2020)

\bibitem{Weiss2021-SA}
Weiss, M., Chakraborty, R., Tonella, P.: A review and refinement of surprise
  adequacy.
\newblock In: 2021 IEEE/ACM Third International Workshop on Deep Learning for
  Testing and Testing for Deep Learning (DeepTest), pp. 17--24. IEEE (2021)

\bibitem{Weiss2021}
Weiss, M., Tonella, P.: Fail-safe execution of deep learning based systems
  through uncertainty monitoring.
\newblock In: 2021 IEEE 14th International Conference on Software Testing,
  Validation and Verification (ICST). IEEE (2021)

\bibitem{Weiss2021Uwiz}
Weiss, M., Tonella, P.: Uncertainty-wizard: Fast and user-friendly neural
  network uncertainty quantification.
\newblock In: 2021 14th IEEE Conference on Software Testing, Verification and
  Validation (ICST), pp. 436--441 (2021).
\newblock \doi{10.1109/ICST49551.2021.00056}

\bibitem{weiss2022simpleTip}
Weiss, M., Tonella, P.: Simple techniques work surprisingly well for neural
  network test prioritization and active learning (replicability study).
\newblock In: Proceedings of the 31st ACM SIGSOFT International Symposium on
  Software Testing and Analysis, ISSTA 2022, p. 139–150. Association for
  Computing Machinery, New York, NY, USA (2022).
\newblock \doi{10.1145/3533767.3534375}.
\newblock \urlprefix\url{https://arxiv.org/abs/2205.00664}

\bibitem{Weiss2022Stvr}
Weiss, M., Tonella, P.: Uncertainty quantification for deep neural networks: An
  empirical comparison and usage guidelines.
\newblock Software Testing, Verification and Reliability  (Forthcoming)

\bibitem{wintersberger2021sdc}
Wintersberger, P., Janotta, F., Peintner, J., L{\"o}cken, A., Riener, A.:
  Evaluating feedback requirements for trust calibration in automated vehicles.
\newblock it-Information Technology \textbf{63}(2), 111--122 (2021)

\bibitem{Xiao2017}
Xiao, H., Rasul, K., Vollgraf, R.: Fashion-mnist: a novel image dataset for
  benchmarking machine learning algorithms (2017)

\bibitem{Xie2019}
Xie, X., Ma, L., Juefei-Xu, F., Xue, M., Chen, H., Liu, Y., Zhao, J., Li, B.,
  Yin, J., See, S.: Deephunter: a coverage-guided fuzz testing framework for
  deep neural networks.
\newblock In: Proceedings of the 28th ACM SIGSOFT International Symposium on
  Software Testing and Analysis, pp. 146--157 (2019)

\bibitem{Zhang2018}
Zhang, M., Zhang, Y., Zhang, L., Liu, C., Khurshid, S.: Deeproad: Gan-based
  metamorphic testing and input validation framework for autonomous driving
  systems.
\newblock In: Proceedings of the 33rd ACM/IEEE International Conference on
  Automated Software Engineering, ASE 2018, pp. 132--142. ACM, New York, NY,
  USA (2018).
\newblock \doi{10.1145/3238147.3238187}.
\newblock \urlprefix\url{http://doi.acm.org/10.1145/3238147.3238187}

\bibitem{Zhang2020}
Zhang, X., Xie, X., Ma, L., Du, X., Hu, Q., Liu, Y., Zhao, J., Sun, M.: Towards
  characterizing adversarial defects of deep learning software from the lens of
  uncertainty.
\newblock In: Proceedings of 42nd International Conference on Software
  Engineering. ACM (2020)

\end{thebibliography}

\newpage
\appendix

\section{Top-Pair Accuracy}
\label{appendix:top-pair}

Above, we defined \emph{Top-Pair Accuracy} as \emph{"the percentage of inputs for which the two most likely predicted classes equal the two true classes between which the input is ambiguous"}. 
As such, top-pair accuracy can only be computed on a dataset of truly ambiguous samples, where for every sample two classes have strictly higher likelihood in the probabilistic label than all other classes. In our ambiguous datasets, where for every sample exactly two classes have nonzero probability, this is naturally given. 

\paragraph{Example} In the following, we provide a specific example showing how Top-Pair Accuracy is calculated. Consider \autoref{tab:top_pair_example}, which shows probabilistic labels and softmax predictions for a dataset with 5 classes and 7 samples. 
Here, we extract the \emph{label top-pair}, i.e., the unordered pair consisting of the two classes with the highest probability labels (in our dataset, these are just the classes with nonzero probability). Then, we do the same with the model predictions, where the \emph{predicted top-pair} consists of the two classes with the highest predicted likelihood. A sample is considered matching if and only if the \emph{label top-pair} equals the \emph{predicted top-pair}. The top-pair accuracy is then computed as the share of matching samples, which, in our example is $\frac{5}{7}=0.714 $.

\newcolumntype{x}[0]{>{\centering\arraybackslash}p{0.25cm}}
\begin{table}[]
\centering
\begin{tabular}{@{}l xxxxxcxxxxxcc@{}}

\toprule
         & \multicolumn{5}{c}{Probabilistic Label}                                                                                                                & \begin{tabular}[c]{@{}c@{}}Label \\ Top-Pair\end{tabular} & \multicolumn{5}{c}{Softmax Predictions}                                                                                                                 & \begin{tabular}[c]{@{}c@{}}Pred.\\ Top-Pair\end{tabular} & Match \\ 
\# & p(0)                         & p(1)                        & p(2)                         & p(3)                         & p(4)                        &                                                           & p(0)                         & p(1)                         & p(2)                        & p(3)                         & p(4)                         &                                                               &       \\\midrule
0        & 0                            & \cellcolor[HTML]{C0C0C0}.4 & 0                            & \cellcolor[HTML]{C0C0C0}.6  & 0                           & \{1,3\}                                                   & .1                          & \cellcolor[HTML]{C0C0C0}.45 & .05                        & \cellcolor[HTML]{C0C0C0}.25 & .15                         & \{1,3\}                                                       & \checkmark  \\
1        & \cellcolor[HTML]{C0C0C0}.45 & 0                           & \cellcolor[HTML]{C0C0C0}.55 & 0                            & 0                           & \{0,2\}                                                   & \cellcolor[HTML]{C0C0C0}.4  & \cellcolor[HTML]{C0C0C0}.45 & .1                         & .02                         & .03                         & \{0,1\}                                                       & x \\
2        & 0                            & \cellcolor[HTML]{C0C0C0}.3 & 0                            & .7                          & 0                           & \{1,3\}                                                   & .03                         & \cellcolor[HTML]{C0C0C0}.6  & \cellcolor[HTML]{C0C0C0}.2 & .1                          & .07                         & \{1,2\}                                                       & x \\
3        & \cellcolor[HTML]{C0C0C0}.35 & 0                           & 0                            & \cellcolor[HTML]{C0C0C0}.65 & 0                           & \{0,3\}                                                   & \cellcolor[HTML]{C0C0C0}.45 & .05                         & .1                         & \cellcolor[HTML]{C0C0C0}.35 & .05                         & \{0,3\}                                                       & \checkmark  \\
4        & 0                            & 0                           & \cellcolor[HTML]{C0C0C0}.5  & 0                            & \cellcolor[HTML]{C0C0C0}.5 & \{2,4\}                                                   & .06                         & .07                         & \cellcolor[HTML]{C0C0C0}.3 & .2                          & \cellcolor[HTML]{C0C0C0}.37 & \{2,4\}                                                       & \checkmark  \\
5        & \cellcolor[HTML]{C0C0C0}.2  & 0                           & 0                            & \cellcolor[HTML]{C0C0C0}.8  & 0                           & \{0,3\}                                                   & \cellcolor[HTML]{C0C0C0}.3  & .03                         & .02                        & \cellcolor[HTML]{C0C0C0}.6  & .05                         & \{0,3\}                                                       & \checkmark  \\
6        & 0                            & \cellcolor[HTML]{C0C0C0}.4 & 0                            & 0                            & \cellcolor[HTML]{C0C0C0}.6 & \{1,4\}                                                   & .1                          & \cellcolor[HTML]{C0C0C0}.35 & .06                        & .04                         & \cellcolor[HTML]{C0C0C0}.45 & \{1,4\}                                                       & \checkmark  \\ \bottomrule
\end{tabular}
\caption{Example for Top-Pair Accuracy Calculation.}
\label{tab:top_pair_example}
\end{table}

\section{DNN-Architecture Specific Results of the Ambiguity Evaluation}
\label{appendix:ambiguity-eval}
This section provides the results which are presented in ~\autoref{tab:ambiguity_res} in an aggregated form for all four used DNN architectures individually. Specifically,
~\autoref{tab:ambiguity_res_convnn} shows the results of the simple convolutional DNN,
~\autoref{tab:ambiguity_res_fullyconnectednn} shows the results of a fully connected DNN,
~\autoref{tab:ambiguity_res_densenet} shows the results with the Densenet architecture~\cite{Huang2017Densenet},
and ~\autoref{tab:ambiguity_res_resnet50} shows the results with the Resnet50 architecture~\cite{He2016Resnet}.
Overall, the results between the four architectures are comparably similar, except for the fully connected DNN which is generally the weakest architecture and thus achieves lower accuracies (but still shows the overall tendencies discussed in ~\autoref{sec:quantitative_evaluation} when assessing the quality of our ambiguous data).

\begin{table}
\centering
\begin{tabular}{@{}llcccc@{}}
\textbf{Training Set}            & \textbf{Test Set} & \textbf{Top-1 Acc} & \textbf{Top-2 Acc} & \textbf{Top-Pair Acc} & \textbf{Entropy} \\\midrule

\multicolumn{6}{c}{\textit{Our dataset for Fashion MNIST}}      \vspace{3px} \\

\multirow{2}{*}{mixed-ambiguous}   & ambiguous  & 0.52 & 0.95 & 0.89 & 1.56\\
                                         & nominal    & 0.90 & 0.97 & n.a. & 0.38\\
\multirow{2}{*}{clean}             & ambiguous  & 0.32 & 0.51 & 0.17 & 1.51\\
                                         & nominal    & 0.90 & 0.98 & n.a. & 0.39\\
    
\midrule
\multicolumn{6}{c}{\textit{Our dataset for MNIST}}      \vspace{3px} \\

\multirow{2}{*}{mixed-ambiguous}   & ambiguous  & 0.54 & 0.99 & 0.98 & 1.48\\
                                         & nominal    & 0.99 & 1.00 & n.a. & 0.06\\
\multirow{2}{*}{clean}             & ambiguous  & 0.47 & 0.73 & 0.46 & 1.07\\
                                         & nominal    & 0.99 & 1.00 & n.a. & 0.04\\
    
\midrule
\multicolumn{6}{c}{\textit{Baseline for mnist: AmbiguousMNIST by Mukhoti et. al.~\cite{Mukhoti2021}}}      \vspace{3px} \\

\multirow{2}{*}{mixed-ambiguous}   & ambiguous  & 0.77 & 0.93 & not calculable & 0.87\\
                                         & nominal    & 0.99 & 1.00 & n.a. & 0.04\\
\multirow{2}{*}{clean}             & ambiguous  & 0.77 & 0.93 & not calculable & 0.81\\
                                         & nominal    & 0.99 & 1.00 & n.a. & 0.03\\

\bottomrule
\end{tabular}
\caption{Evaluation of Ambiguity (Conv. NN)}
\label{tab:ambiguity_res_convnn}
\end{table}

\begin{table}
\centering
\begin{tabular}{@{}llcccc@{}}
\textbf{Training Set}            & \textbf{Test Set} & \textbf{Top-1 Acc} & \textbf{Top-2 Acc} & \textbf{Top-Pair Acc} & \textbf{Entropy} \\\midrule

\multicolumn{6}{c}{\textit{Our dataset for Fashion MNIST}}      \vspace{3px} \\

\multirow{2}{*}{mixed-ambiguous}   & ambiguous  & 0.49 & 0.82 & 0.63 & 1.78\\
                                         & nominal    & 0.82 & 0.93 & n.a. & 0.69\\
\multirow{2}{*}{clean}             & ambiguous  & 0.32 & 0.50 & 0.13 & 1.23\\
                                         & nominal    & 0.83 & 0.94 & n.a. & 0.57\\
    
\midrule
\multicolumn{6}{c}{\textit{Our dataset for MNIST}}      \vspace{3px} \\

\multirow{2}{*}{mixed-ambiguous}   & ambiguous  & 0.53 & 0.94 & 0.86 & 1.45\\
                                         & nominal    & 0.91 & 0.96 & n.a. & 0.49\\
\multirow{2}{*}{clean}             & ambiguous  & 0.45 & 0.70 & 0.43 & 1.10\\
                                         & nominal    & 0.92 & 0.97 & n.a. & 0.34\\
    
\midrule
\multicolumn{6}{c}{\textit{Baseline for mnist: AmbiguousMNIST by Mukhoti et. al.~\cite{Mukhoti2021}}}      \vspace{3px} \\

\multirow{2}{*}{mixed-ambiguous}   & ambiguous  & 0.63 & 0.86 & not calculable & 1.31\\
                                         & nominal    & 0.92 & 0.97 & n.a. & 0.37\\
\multirow{2}{*}{clean}             & ambiguous  & 0.56 & 0.80 & not calculable & 1.16\\
                                         & nominal    & 0.92 & 0.97 & n.a. & 0.36\\

\bottomrule
\end{tabular}
\caption{Evaluation of Ambiguity (Fully Connected NN.)}
\label{tab:ambiguity_res_fullyconnectednn}
\end{table}

\begin{table}
\centering
\begin{tabular}{@{}llcccc@{}}
\textbf{Training Set}            & \textbf{Test Set} & \textbf{Top-1 Acc} & \textbf{Top-2 Acc} & \textbf{Top-Pair Acc} & \textbf{Entropy} \\\midrule

\multicolumn{6}{c}{\textit{Our dataset for Fashion MNIST}}      \vspace{3px} \\

\multirow{2}{*}{mixed-ambiguous}   & ambiguous  & 0.52 & 0.99 & 0.98 & 0.98\\
                                         & nominal    & 0.90 & 0.98 & n.a. & 0.22\\
\multirow{2}{*}{clean}             & ambiguous  & 0.28 & 0.44 & 0.12 & 0.71\\
                                         & nominal    & 0.89 & 0.97 & n.a. & 0.23\\
    
\midrule
\multicolumn{6}{c}{\textit{Our dataset for MNIST}}      \vspace{3px} \\

\multirow{2}{*}{mixed-ambiguous}   & ambiguous  & 0.53 & 1.00 & 1.00 & 0.96\\
                                         & nominal    & 0.99 & 1.00 & n.a. & 0.02\\
\multirow{2}{*}{clean}             & ambiguous  & 0.35 & 0.53 & 0.20 & 0.31\\
                                         & nominal    & 0.99 & 1.00 & n.a. & 0.03\\
    
\midrule
\multicolumn{6}{c}{\textit{Baseline for mnist: AmbiguousMNIST by Mukhoti et. al.~\cite{Mukhoti2021}}}      \vspace{3px} \\

\multirow{2}{*}{mixed-ambiguous}   & ambiguous  & 0.75 & 0.93 & not calculable & 0.67\\
                                         & nominal    & 0.99 & 1.00 & n.a. & 0.03\\
\multirow{2}{*}{clean}             & ambiguous  & 0.58 & 0.83 & not calculable & 0.34\\
                                         & nominal    & 0.98 & 1.00 & n.a. & 0.04\\

\bottomrule
\end{tabular}
\caption{Evaluation of Ambiguity (Densenet)}
\label{tab:ambiguity_res_densenet}
\end{table}

\begin{table}
\centering
\begin{tabular}{@{}llcccc@{}}
\textbf{Training Set}            & \textbf{Test Set} & \textbf{Top-1 Acc} & \textbf{Top-2 Acc} & \textbf{Top-Pair Acc} & \textbf{Entropy} \\\midrule

\multicolumn{6}{c}{\textit{Our dataset for Fashion MNIST}}      \vspace{3px} \\

\multirow{2}{*}{mixed-ambiguous}   & ambiguous  & 0.51 & 0.99 & 0.97 & 1.01\\
                                         & nominal    & 0.92 & 0.98 & n.a. & 0.12\\
\multirow{2}{*}{clean}             & ambiguous  & 0.36 & 0.50 & 0.12 & 0.76\\
                                         & nominal    & 0.92 & 0.98 & n.a. & 0.15\\
    
\midrule
\multicolumn{6}{c}{\textit{Our dataset for MNIST}}      \vspace{3px} \\

\multirow{2}{*}{mixed-ambiguous}   & ambiguous  & 0.53 & 0.99 & 0.99 & 0.99\\
                                         & nominal    & 0.99 & 1.00 & n.a. & 0.02\\
\multirow{2}{*}{clean}             & ambiguous  & 0.42 & 0.59 & 0.21 & 0.43\\
                                         & nominal    & 0.99 & 1.00 & n.a. & 0.02\\
    
\midrule
\multicolumn{6}{c}{\textit{Baseline for mnist: AmbiguousMNIST by Mukhoti et. al.~\cite{Mukhoti2021}}}      \vspace{3px} \\

\multirow{2}{*}{mixed-ambiguous}   & ambiguous  & 0.76 & 0.92 & not calculable & 0.67\\
                                         & nominal    & 0.99 & 1.00 & n.a. & 0.03\\
\multirow{2}{*}{clean}             & ambiguous  & 0.67 & 0.83 & not calculable & 0.41\\
                                         & nominal    & 0.99 & 1.00 & n.a. & 0.02\\

\bottomrule
\end{tabular}
\caption{Evaluation of Ambiguity (Resnet50)}
\label{tab:ambiguity_res_resnet50}
\end{table}

\section{DNN-Architecture Specific Comparison of Supervisors}
\label{appendix:supervisor-eval}
This section provides information of the performance of the different supervisors for each of the four supervised architectures (Tables~\ref{tab:auc_roc_simplecnn},~\ref{tab:auc_roc_fullyconnectednet},~\ref{tab:auc_roc_densenet}~and~\ref{tab:auc_roc_resnet}). For each of these architectures, five models were trained to account for the randomness faced during training. The corresponding standard deviations are also reported in the tables.

    \newcolumntype{Y}{>{\centering\arraybackslash}X}
    
    \begin{table*}[t]\small
    \begin{tabularx}{\linewidth}{l|YYYY|YYYY}
\toprule
{} & \multicolumn{4}{c}{mnist} & \multicolumn{4}{c}{fmnist} \\
{} &                       amb. &                     adv. &                       corr. &                          inv. &                       amb. &                     adv. &                       corr. &                         inv. \\
\midrule

\multicolumn{9}{l}{\small \textit{Plain Softmax Supervisors}}\vspace{2px}\\Max. SM.       &  .96 +- \newline\textit{.00} &  .79 +- \newline\textit{.01} &  .78 +- \newline\textit{.00} &   .79 +- \newline\textit{.00} &  .91 +- \newline\textit{.01} &  .61 +- \newline\textit{.02} &  .71 +- \newline\textit{.01} &  .73 +- \newline\textit{.02} \\
PCS                &  .96 +- \newline\textit{.00} &  .79 +- \newline\textit{.01} &  .78 +- \newline\textit{.00} &   .79 +- \newline\textit{.00} &  .91 +- \newline\textit{.01} &  .61 +- \newline\textit{.02} &  .70 +- \newline\textit{.01} &  .72 +- \newline\textit{.02} \\
SM. Ent.    &  .97 +- \newline\textit{.00} &  .79 +- \newline\textit{.01} &  .78 +- \newline\textit{.00} &   .79 +- \newline\textit{.00} &  .92 +- \newline\textit{.00} &  .61 +- \newline\textit{.02} &  .72 +- \newline\textit{.01} &  .74 +- \newline\textit{.02} \\
DeepGini           &  .96 +- \newline\textit{.00} &  .79 +- \newline\textit{.01} &  .78 +- \newline\textit{.00} &   .79 +- \newline\textit{.00} &  .92 +- \newline\textit{.00} &  .61 +- \newline\textit{.02} &  .71 +- \newline\textit{.01} &  .73 +- \newline\textit{.02} \vspace{2px}\\
\multicolumn{9}{l}{\small \textit{Monte-Carlo Dropout Supervisors (Softmax-based, except for VR)}}\vspace{2px}\\VR    &  .79 +- \newline\textit{.00} &  .69 +- \newline\textit{.01} &  .65 +- \newline\textit{.01} &   .72 +- \newline\textit{.02} &  .76 +- \newline\textit{.01} &  .62 +- \newline\textit{.01} &  .66 +- \newline\textit{.01} &  .72 +- \newline\textit{.01} \\
MS    &  .96 +- \newline\textit{.00} &  .79 +- \newline\textit{.01} &  .80 +- \newline\textit{.00} &   .80 +- \newline\textit{.01} &  .91 +- \newline\textit{.01} &  .61 +- \newline\textit{.02} &  .73 +- \newline\textit{.01} &  .77 +- \newline\textit{.02} \\
MI    &  .87 +- \newline\textit{.00} &  .78 +- \newline\textit{.01} &  .81 +- \newline\textit{.00} &   .83 +- \newline\textit{.01} &  .73 +- \newline\textit{.01} &  .61 +- \newline\textit{.02} &  .78 +- \newline\textit{.01} &  .86 +- \newline\textit{.02} \\
PE    &  .96 +- \newline\textit{.00} &  .79 +- \newline\textit{.01} &  .80 +- \newline\textit{.00} &   .80 +- \newline\textit{.00} &  .91 +- \newline\textit{.00} &  .61 +- \newline\textit{.02} &  .74 +- \newline\textit{.01} &  .79 +- \newline\textit{.02} \vspace{2px}\\
\multicolumn{9}{l}{\small \textit{Deep Ensemble Supervisors (Softmax-based)}}\vspace{2px}\\MS &  .97 +- \newline\textit{.00} &                        n.a. &  .84 +- \newline\textit{.00} &   .85 +- \newline\textit{.00} &  .90 +- \newline\textit{.00} &                        n.a. &  .75 +- \newline\textit{.00} &  .64 +- \newline\textit{.01} \\
MI &  .84 +- \newline\textit{.00} &                        n.a. &  .84 +- \newline\textit{.00} &   .88 +- \newline\textit{.00} &  .57 +- \newline\textit{.01} &                        n.a. &  .76 +- \newline\textit{.01} &  .70 +- \newline\textit{.01} \\
PE &  .97 +- \newline\textit{.00} &                        n.a. &  .83 +- \newline\textit{.00} &   .84 +- \newline\textit{.00} &  .89 +- \newline\textit{.00} &                        n.a. &  .77 +- \newline\textit{.00} &  .66 +- \newline\textit{.01} \vspace{2px}\\
\multicolumn{9}{l}{\small \textit{Other Supervisors}}\vspace{2px}\\Dissector          &  .95 +- \newline\textit{.00} &  .79 +- \newline\textit{.01} &  .76 +- \newline\textit{.00} &   .79 +- \newline\textit{.01} &  .88 +- \newline\textit{.00} &  .68 +- \newline\textit{.01} &  .72 +- \newline\textit{.00} &  .75 +- \newline\textit{.01} \\
DSA                &  .48 +- \newline\textit{.01} &  .93 +- \newline\textit{.00} &  .87 +- \newline\textit{.00} &   .98 +- \newline\textit{.00} &  .31 +- \newline\textit{.01} &  .85 +- \newline\textit{.01} &  .85 +- \newline\textit{.00} &  .90 +- \newline\textit{.00} \\
LSA                &  .17 +- \newline\textit{.01} &  .78 +- \newline\textit{.02} &  .73 +- \newline\textit{.01} &   .77 +- \newline\textit{.03} &  .16 +- \newline\textit{.00} &  .75 +- \newline\textit{.01} &  .74 +- \newline\textit{.01} &  .86 +- \newline\textit{.00} \\
MDSA               &  .31 +- \newline\textit{.01} &  .94 +- \newline\textit{.01} &  .87 +- \newline\textit{.00} &   .98 +- \newline\textit{.00} &  .32 +- \newline\textit{.01} &  .86 +- \newline\textit{.01} &  .83 +- \newline\textit{.01} &  .95 +- \newline\textit{.00} \\
Autoenc.        &  .62 +- \newline\textit{.00} &  .95 +- \newline\textit{.00} &  .84 +- \newline\textit{.00} &  1.00 +- \newline\textit{.00} &  .53 +- \newline\textit{.00} &  .80 +- \newline\textit{.01} &  .77 +- \newline\textit{.00} &  .49 +- \newline\textit{.01} \\
\bottomrule
\end{tabularx}
\caption{Supervisor's performance at discriminating nominal from high-uncertainty inputs (AUC-ROC), for the SimpleCnn architecture.}
\label{tab:auc_roc_simplecnn}
\end{table*}

    \newcolumntype{Y}{>{\centering\arraybackslash}X}
    
    \begin{table*}[t]\small
    \begin{tabularx}{\linewidth}{l|YYYY|YYYY}
\toprule
{} & \multicolumn{4}{c}{mnist} & \multicolumn{4}{c}{fmnist} \\
{} &                       amb. &                     adv. &                       corr. &                          inv. &                       amb. &                     adv. &                       corr. &                         inv. \\
\midrule

\multicolumn{9}{l}{\small \textit{Plain Softmax Supervisors}}\vspace{2px}\\Max. SM.       &  .96 +- \newline\textit{.01} &  .79 +- \newline\textit{.01} &  .78 +- \newline\textit{.04} &   .79 +- \newline\textit{.04} &  .91 +- \newline\textit{.03} &  .61 +- \newline\textit{.01} &  .71 +- \newline\textit{.02} &  .73 +- \newline\textit{.02} \\
PCS                &  .96 +- \newline\textit{.01} &  .79 +- \newline\textit{.01} &  .78 +- \newline\textit{.04} &   .79 +- \newline\textit{.04} &  .91 +- \newline\textit{.02} &  .61 +- \newline\textit{.01} &  .70 +- \newline\textit{.02} &  .72 +- \newline\textit{.02} \\
SM. Ent.    &  .97 +- \newline\textit{.01} &  .79 +- \newline\textit{.01} &  .78 +- \newline\textit{.03} &   .79 +- \newline\textit{.04} &  .92 +- \newline\textit{.03} &  .61 +- \newline\textit{.01} &  .72 +- \newline\textit{.02} &  .74 +- \newline\textit{.02} \\
DeepGini           &  .96 +- \newline\textit{.01} &  .79 +- \newline\textit{.01} &  .78 +- \newline\textit{.04} &   .79 +- \newline\textit{.04} &  .92 +- \newline\textit{.03} &  .61 +- \newline\textit{.01} &  .71 +- \newline\textit{.02} &  .73 +- \newline\textit{.02} \vspace{2px}\\
\multicolumn{9}{l}{\small \textit{Monte-Carlo Dropout Supervisors (Softmax-based, except for VR)}}\vspace{2px}\\VR    &  .79 +- \newline\textit{.01} &  .69 +- \newline\textit{.00} &  .65 +- \newline\textit{.02} &   .72 +- \newline\textit{.04} &  .76 +- \newline\textit{.02} &  .62 +- \newline\textit{.01} &  .66 +- \newline\textit{.02} &  .72 +- \newline\textit{.01} \\
MS    &  .96 +- \newline\textit{.01} &  .79 +- \newline\textit{.01} &  .80 +- \newline\textit{.04} &   .80 +- \newline\textit{.06} &  .91 +- \newline\textit{.03} &  .61 +- \newline\textit{.01} &  .73 +- \newline\textit{.02} &  .77 +- \newline\textit{.02} \\
MI    &  .87 +- \newline\textit{.01} &  .78 +- \newline\textit{.01} &  .81 +- \newline\textit{.02} &   .83 +- \newline\textit{.10} &  .73 +- \newline\textit{.02} &  .61 +- \newline\textit{.01} &  .78 +- \newline\textit{.02} &  .86 +- \newline\textit{.02} \\
PE    &  .96 +- \newline\textit{.01} &  .79 +- \newline\textit{.01} &  .80 +- \newline\textit{.03} &   .80 +- \newline\textit{.06} &  .91 +- \newline\textit{.03} &  .61 +- \newline\textit{.01} &  .74 +- \newline\textit{.02} &  .79 +- \newline\textit{.02} \vspace{2px}\\
\multicolumn{9}{l}{\small \textit{Deep Ensemble Supervisors (Softmax-based)}}\vspace{2px}\\MS &  .97 +- \newline\textit{.00} &                        n.a. &  .84 +- \newline\textit{.00} &   .85 +- \newline\textit{.03} &  .90 +- \newline\textit{.00} &                        n.a. &  .75 +- \newline\textit{.00} &  .64 +- \newline\textit{.00} \\
MI &  .84 +- \newline\textit{.01} &                        n.a. &  .84 +- \newline\textit{.00} &   .88 +- \newline\textit{.05} &  .57 +- \newline\textit{.03} &                        n.a. &  .76 +- \newline\textit{.01} &  .70 +- \newline\textit{.00} \\
PE &  .97 +- \newline\textit{.00} &                        n.a. &  .83 +- \newline\textit{.00} &   .84 +- \newline\textit{.03} &  .89 +- \newline\textit{.00} &                        n.a. &  .77 +- \newline\textit{.00} &  .66 +- \newline\textit{.00} \vspace{2px}\\
\multicolumn{9}{l}{\small \textit{Other Supervisors}}\vspace{2px}\\Dissector          &  .95 +- \newline\textit{.01} &  .79 +- \newline\textit{.00} &  .76 +- \newline\textit{.05} &   .79 +- \newline\textit{.12} &  .88 +- \newline\textit{.01} &  .68 +- \newline\textit{.01} &  .72 +- \newline\textit{.04} &  .75 +- \newline\textit{.02} \\
DSA                &  .48 +- \newline\textit{.01} &  .93 +- \newline\textit{.01} &  .87 +- \newline\textit{.00} &   .98 +- \newline\textit{.00} &  .31 +- \newline\textit{.02} &  .85 +- \newline\textit{.02} &  .85 +- \newline\textit{.00} &  .90 +- \newline\textit{.01} \\
LSA                &  .17 +- \newline\textit{.00} &  .78 +- \newline\textit{.01} &  .73 +- \newline\textit{.00} &   .77 +- \newline\textit{.00} &  .16 +- \newline\textit{.01} &  .75 +- \newline\textit{.01} &  .74 +- \newline\textit{.00} &  .86 +- \newline\textit{.00} \\
MDSA               &  .31 +- \newline\textit{.00} &  .94 +- \newline\textit{.01} &  .87 +- \newline\textit{.00} &   .98 +- \newline\textit{.00} &  .32 +- \newline\textit{.00} &  .86 +- \newline\textit{.01} &  .83 +- \newline\textit{.00} &  .95 +- \newline\textit{.00} \\
Autoenc.        &  .62 +- \newline\textit{.01} &  .95 +- \newline\textit{.01} &  .84 +- \newline\textit{.00} &  1.00 +- \newline\textit{.00} &  .53 +- \newline\textit{.01} &  .80 +- \newline\textit{.04} &  .77 +- \newline\textit{.01} &  .49 +- \newline\textit{.08} \\
\bottomrule
\end{tabularx}
\caption{Supervisor's performance at discriminating nominal from high-uncertainty inputs (AUC-ROC), for the FullyConnectedNet architecture.}
\label{tab:auc_roc_fullyconnectednet}
\end{table*}

    \newcolumntype{Y}{>{\centering\arraybackslash}X}
    
    \begin{table*}[t]\small
    \begin{tabularx}{\linewidth}{l|YYYY|YYYY}
\toprule
{} & \multicolumn{4}{c}{mnist} & \multicolumn{4}{c}{fmnist} \\
{} &                       amb. &                     adv. &                       corr. &                          inv. &                       amb. &                     adv. &                       corr. &                         inv. \\
\midrule

\multicolumn{9}{l}{\small \textit{Plain Softmax Supervisors}}\vspace{2px}\\Max. SM.       &  .96 +- \newline\textit{.00} &  .79 +- \newline\textit{.02} &  .78 +- \newline\textit{.02} &   .79 +- \newline\textit{.01} &  .91 +- \newline\textit{.00} &  .61 +- \newline\textit{.02} &  .71 +- \newline\textit{.02} &  .73 +- \newline\textit{.08} \\
PCS                &  .96 +- \newline\textit{.00} &  .79 +- \newline\textit{.02} &  .78 +- \newline\textit{.02} &   .79 +- \newline\textit{.01} &  .91 +- \newline\textit{.01} &  .61 +- \newline\textit{.02} &  .70 +- \newline\textit{.02} &  .72 +- \newline\textit{.08} \\
SM. Ent.    &  .97 +- \newline\textit{.01} &  .79 +- \newline\textit{.02} &  .78 +- \newline\textit{.02} &   .79 +- \newline\textit{.01} &  .92 +- \newline\textit{.00} &  .61 +- \newline\textit{.02} &  .72 +- \newline\textit{.02} &  .74 +- \newline\textit{.08} \\
DeepGini           &  .96 +- \newline\textit{.00} &  .79 +- \newline\textit{.02} &  .78 +- \newline\textit{.02} &   .79 +- \newline\textit{.01} &  .92 +- \newline\textit{.00} &  .61 +- \newline\textit{.02} &  .71 +- \newline\textit{.02} &  .73 +- \newline\textit{.08} \vspace{2px}\\
\multicolumn{9}{l}{\small \textit{Monte-Carlo Dropout Supervisors (Softmax-based, except for VR)}}\vspace{2px}\\VR    &  .79 +- \newline\textit{.02} &  .69 +- \newline\textit{.01} &  .65 +- \newline\textit{.02} &   .72 +- \newline\textit{.03} &  .76 +- \newline\textit{.02} &  .62 +- \newline\textit{.02} &  .66 +- \newline\textit{.01} &  .72 +- \newline\textit{.07} \\
MS    &  .96 +- \newline\textit{.00} &  .79 +- \newline\textit{.02} &  .80 +- \newline\textit{.02} &   .80 +- \newline\textit{.01} &  .91 +- \newline\textit{.01} &  .61 +- \newline\textit{.02} &  .73 +- \newline\textit{.02} &  .77 +- \newline\textit{.08} \\
MI    &  .87 +- \newline\textit{.05} &  .78 +- \newline\textit{.03} &  .81 +- \newline\textit{.02} &   .83 +- \newline\textit{.01} &  .73 +- \newline\textit{.02} &  .61 +- \newline\textit{.02} &  .78 +- \newline\textit{.02} &  .86 +- \newline\textit{.07} \\
PE    &  .96 +- \newline\textit{.01} &  .79 +- \newline\textit{.02} &  .80 +- \newline\textit{.02} &   .80 +- \newline\textit{.01} &  .91 +- \newline\textit{.00} &  .61 +- \newline\textit{.02} &  .74 +- \newline\textit{.02} &  .79 +- \newline\textit{.09} \vspace{2px}\\
\multicolumn{9}{l}{\small \textit{Deep Ensemble Supervisors (Softmax-based)}}\vspace{2px}\\MS &  .97 +- \newline\textit{.00} &                        n.a. &  .84 +- \newline\textit{.00} &   .85 +- \newline\textit{.00} &  .90 +- \newline\textit{.00} &                        n.a. &  .75 +- \newline\textit{.00} &  .64 +- \newline\textit{.02} \\
MI &  .84 +- \newline\textit{.01} &                        n.a. &  .84 +- \newline\textit{.00} &   .88 +- \newline\textit{.00} &  .57 +- \newline\textit{.01} &                        n.a. &  .76 +- \newline\textit{.01} &  .70 +- \newline\textit{.02} \\
PE &  .97 +- \newline\textit{.00} &                        n.a. &  .83 +- \newline\textit{.00} &   .84 +- \newline\textit{.00} &  .89 +- \newline\textit{.00} &                        n.a. &  .77 +- \newline\textit{.01} &  .66 +- \newline\textit{.02} \vspace{2px}\\
\multicolumn{9}{l}{\small \textit{Other Supervisors}}\vspace{2px}\\Dissector          &  .95 +- \newline\textit{.01} &  .79 +- \newline\textit{.04} &  .76 +- \newline\textit{.02} &   .79 +- \newline\textit{.04} &  .88 +- \newline\textit{.01} &  .68 +- \newline\textit{.01} &  .72 +- \newline\textit{.01} &  .75 +- \newline\textit{.08} \\
DSA                &  .48 +- \newline\textit{.07} &  .93 +- \newline\textit{.01} &  .87 +- \newline\textit{.01} &   .98 +- \newline\textit{.00} &  .31 +- \newline\textit{.01} &  .85 +- \newline\textit{.02} &  .85 +- \newline\textit{.01} &  .90 +- \newline\textit{.03} \\
LSA                &  .17 +- \newline\textit{.02} &  .78 +- \newline\textit{.00} &  .73 +- \newline\textit{.00} &   .77 +- \newline\textit{.00} &  .16 +- \newline\textit{.01} &  .75 +- \newline\textit{.01} &  .74 +- \newline\textit{.00} &  .86 +- \newline\textit{.00} \\
MDSA               &  .31 +- \newline\textit{.04} &  .94 +- \newline\textit{.00} &  .87 +- \newline\textit{.00} &   .98 +- \newline\textit{.01} &  .32 +- \newline\textit{.04} &  .86 +- \newline\textit{.01} &  .83 +- \newline\textit{.00} &  .95 +- \newline\textit{.00} \\
Autoenc.        &  .62 +- \newline\textit{.03} &  .95 +- \newline\textit{.01} &  .84 +- \newline\textit{.05} &  1.00 +- \newline\textit{.01} &  .53 +- \newline\textit{.01} &  .80 +- \newline\textit{.03} &  .77 +- \newline\textit{.01} &  .49 +- \newline\textit{.07} \\
\bottomrule
\end{tabularx}
\caption{Supervisor's performance at discriminating nominal from high-uncertainty inputs (AUC-ROC), for the Densenet architecture.}
\label{tab:auc_roc_densenet}
\end{table*}

    \newcolumntype{Y}{>{\centering\arraybackslash}X}
    
    \begin{table*}[t]\small
    \begin{tabularx}{\linewidth}{l|YYYY|YYYY}
\toprule
{} & \multicolumn{4}{c}{mnist} & \multicolumn{4}{c}{fmnist} \\
{} &                       amb. &                     adv. &                       corr. &                          inv. &                       amb. &                     adv. &                       corr. &                         inv. \\
\midrule

\multicolumn{9}{l}{\small \textit{Plain Softmax Supervisors}}\vspace{2px}\\Max. SM.       &  .96 +- \newline\textit{.00} &  .79 +- \newline\textit{.04} &  .78 +- \newline\textit{.01} &   .79 +- \newline\textit{.03} &  .91 +- \newline\textit{.00} &  .61 +- \newline\textit{.01} &  .71 +- \newline\textit{.02} &  .73 +- \newline\textit{.06} \\
PCS                &  .96 +- \newline\textit{.00} &  .79 +- \newline\textit{.04} &  .78 +- \newline\textit{.01} &   .79 +- \newline\textit{.03} &  .91 +- \newline\textit{.00} &  .61 +- \newline\textit{.01} &  .70 +- \newline\textit{.02} &  .72 +- \newline\textit{.06} \\
SM. Ent.    &  .97 +- \newline\textit{.00} &  .79 +- \newline\textit{.04} &  .78 +- \newline\textit{.01} &   .79 +- \newline\textit{.03} &  .92 +- \newline\textit{.01} &  .61 +- \newline\textit{.01} &  .72 +- \newline\textit{.02} &  .74 +- \newline\textit{.07} \\
DeepGini           &  .96 +- \newline\textit{.00} &  .79 +- \newline\textit{.04} &  .78 +- \newline\textit{.01} &   .79 +- \newline\textit{.03} &  .92 +- \newline\textit{.01} &  .61 +- \newline\textit{.01} &  .71 +- \newline\textit{.02} &  .73 +- \newline\textit{.06} \vspace{2px}\\
\multicolumn{9}{l}{\small \textit{Monte-Carlo Dropout Supervisors (Softmax-based, except for VR)}}\vspace{2px}\\VR    &  .79 +- \newline\textit{.01} &  .69 +- \newline\textit{.03} &  .65 +- \newline\textit{.01} &   .72 +- \newline\textit{.03} &  .76 +- \newline\textit{.03} &  .62 +- \newline\textit{.01} &  .66 +- \newline\textit{.01} &  .72 +- \newline\textit{.03} \\
MS    &  .96 +- \newline\textit{.00} &  .79 +- \newline\textit{.03} &  .80 +- \newline\textit{.01} &   .80 +- \newline\textit{.03} &  .91 +- \newline\textit{.00} &  .61 +- \newline\textit{.01} &  .73 +- \newline\textit{.02} &  .77 +- \newline\textit{.06} \\
MI    &  .87 +- \newline\textit{.01} &  .78 +- \newline\textit{.03} &  .81 +- \newline\textit{.01} &   .83 +- \newline\textit{.03} &  .73 +- \newline\textit{.02} &  .61 +- \newline\textit{.01} &  .78 +- \newline\textit{.01} &  .86 +- \newline\textit{.06} \\
PE    &  .96 +- \newline\textit{.00} &  .79 +- \newline\textit{.03} &  .80 +- \newline\textit{.01} &   .80 +- \newline\textit{.03} &  .91 +- \newline\textit{.01} &  .61 +- \newline\textit{.01} &  .74 +- \newline\textit{.02} &  .79 +- \newline\textit{.06} \vspace{2px}\\
\multicolumn{9}{l}{\small \textit{Deep Ensemble Supervisors (Softmax-based)}}\vspace{2px}\\MS &  .97 +- \newline\textit{.00} &                        n.a. &  .84 +- \newline\textit{.02} &   .85 +- \newline\textit{.00} &  .90 +- \newline\textit{.00} &                        n.a. &  .75 +- \newline\textit{.00} &  .64 +- \newline\textit{.01} \\
MI &  .84 +- \newline\textit{.12} &                        n.a. &  .84 +- \newline\textit{.03} &   .88 +- \newline\textit{.01} &  .57 +- \newline\textit{.02} &                        n.a. &  .76 +- \newline\textit{.01} &  .70 +- \newline\textit{.01} \\
PE &  .97 +- \newline\textit{.00} &                        n.a. &  .83 +- \newline\textit{.02} &   .84 +- \newline\textit{.00} &  .89 +- \newline\textit{.00} &                        n.a. &  .77 +- \newline\textit{.00} &  .66 +- \newline\textit{.01} \vspace{2px}\\
\multicolumn{9}{l}{\small \textit{Other Supervisors}}\vspace{2px}\\Dissector          &  .95 +- \newline\textit{.00} &  .79 +- \newline\textit{.02} &  .76 +- \newline\textit{.01} &   .79 +- \newline\textit{.01} &  .88 +- \newline\textit{.01} &  .68 +- \newline\textit{.02} &  .72 +- \newline\textit{.02} &  .75 +- \newline\textit{.05} \\
DSA                &  .48 +- \newline\textit{.03} &  .93 +- \newline\textit{.01} &  .87 +- \newline\textit{.01} &   .98 +- \newline\textit{.01} &  .31 +- \newline\textit{.06} &  .85 +- \newline\textit{.01} &  .85 +- \newline\textit{.01} &  .90 +- \newline\textit{.03} \\
LSA                &  .17 +- \newline\textit{.00} &  .78 +- \newline\textit{.00} &  .73 +- \newline\textit{.00} &   .77 +- \newline\textit{.00} &  .16 +- \newline\textit{.00} &  .75 +- \newline\textit{.00} &  .74 +- \newline\textit{.00} &  .86 +- \newline\textit{.00} \\
MDSA               &  .31 +- \newline\textit{.05} &  .94 +- \newline\textit{.02} &  .87 +- \newline\textit{.00} &   .98 +- \newline\textit{.01} &  .32 +- \newline\textit{.05} &  .86 +- \newline\textit{.01} &  .83 +- \newline\textit{.02} &  .95 +- \newline\textit{.01} \\
Autoenc.        &  .62 +- \newline\textit{.00} &  .95 +- \newline\textit{.00} &  .84 +- \newline\textit{.00} &  1.00 +- \newline\textit{.00} &  .53 +- \newline\textit{.00} &  .80 +- \newline\textit{.01} &  .77 +- \newline\textit{.00} &  .49 +- \newline\textit{.01} \\
\bottomrule
\end{tabularx}
\caption{Supervisor's performance at discriminating nominal from high-uncertainty inputs (AUC-ROC), for the Resnet architecture.}
\label{tab:auc_roc_resnet}
\end{table*}

\end{document}